\documentclass[aps,prl,twocolumn,groupedaddress]{revtex4-2}

\usepackage{graphicx}
\usepackage{amsmath}
\usepackage{amssymb}
\usepackage{siunitx}
\usepackage{bm}
\usepackage{hyperref}
\usepackage[capitalise]{cleveref}
\makeatletter
\AtBeginDocument{\let\NAT@space\@empty}
\makeatother
\makeatletter
\def\doauthor#1#2#3{%
  \ignorespaces#1\unskip\@listcomma
  \begingroup
   #3%
  \@if@empty{#2}{\endgroup{}{}}{\endgroup{\comma@space}{}\author@notes{#2}}%
  \space \@listand
}%
\def\author@notes#1{%
  \begingroup
   \let\author@notesep\@empty
   \let\@join\author@notes@join
   #1%
  \endgroup
}%
\def\author@notes@join#1#2#3{%
  \def\author@tempa{#2}%
  \ifx\author@tempa\@empty\else#2\fi
  \author@notesep\frontmatter@footnote{#3}%
  \def\author@notesep{\textsuperscript{,}}%
}%
\let\frontmatter@RRAP@format@orig\frontmatter@RRAP@format
\def\frontmatter@RRAP@format{\vskip\baselineskip\frontmatter@RRAP@format@orig}%
\let\frontmatter@footnote@produce\frontmatter@footnote@produce@footnote
\makeatother

\crefrangelabelformat{equation}{(#3#1#4)--(#5#2#6)}
\crefmultiformat{equation}{Eqs.~(#2#1#3)}{ and~(#2#1#3)}{, (#2#1#3)}{, and~(#2#1#3)}
\crefrangeformat{equation}{Eqs.~(#3#1#4)--(#5#2#6)}
\Crefmultiformat{equation}{Equations~(#2#1#3)}{ and~(#2#1#3)}{, (#2#1#3)}{, and~(#2#1#3)}
\Crefrangeformat{equation}{Equations~(#3#1#4)--(#5#2#6)}
\crefname{section}{Appendix}{Appendices}
\Crefname{section}{Appendix}{Appendices}
\newcommand{\appendixheading}[2]{%
  \refstepcounter{section}\label{#1}%
  \addcontentsline{toc}{section}{Appendix~\thesection: #2}%
  \textit{Appendix~\thesection: #2}---%
}
\usepackage[utf8]{inputenc}
\usepackage[T1]{fontenc}
\usepackage{lmodern}
\hypersetup{
	colorlinks=false,
	pdfborder={0 0 0},
    pdftitle={Optomechanical Limit on Intracavity Intensities from Bulk Mirror Vibrations},
	pdfauthor={Lothar Maisenbacher, Ashwin Singh, Isaac Matteo Pope, Holger Müller},
	pdflang={en-US}
}

\newcommand{\TEMfundamental}{\mathrm{TEM}_{00,(q)}}
\newcommand{\TEMHOM}{\mathrm{TEM}_{10,(q-1)}}
\newcommand{\TEMOHOM}{\mathrm{TEM}_{01,(q-1)}}
\newcommand{\TMS}{\Delta\nu_{01}}

\newcommand{\RS}{R}

\newcommand{\zs}{z_S}

\newcommand{\wzeroopt}{W_{0}}

\newcommand{\wzsopt}{W(\zs)}
\newcommand{\wmech}{W_\mathrm{m}}
\newcommand{\wmechIndex}{W_{\mathrm{m},j}}

\newcommand{\Pcirc}{P_{\mathrm{circ},0}}
\newcommand{\Pthresh}{P_\mathrm{th}}

\newcommand{\PthreshIndex}{P_{\mathrm{th},j}}
\newcommand{\Pthreshzero}{P_{\mathrm{th,res}}}
\newcommand{\Deltaomega}{\Delta\omega_j}
\newcommand{\Deltaomegaeq}{\Delta\omega_{\mathrm{stable\,eq}}}
\newcommand{\DeltaomegaInit}{\Delta\omega_{\mathrm{init}}}
\newcommand{\NA}{\mathrm{NA}}

\newcommand{\m}{\mathrm{m}}
\newcommand{\mIndex}{{\mathrm{m},j}}

\newcommand{\finesse}{\mathcal{F}}
\newcommand{\PIFreq}{f_\mathrm{PI}}
\newcommand{\FSRmech}{\mathrm{FSR}_\mathrm{m}}
\newcommand{\TMSmech}{\mathrm{TMS}_\mathrm{m}}

\newcommand{\phantomeq}[2]{%
  \begingroup
  \renewcommand{\theequation}{#1}%
  \renewcommand{\theHequation}{phantom.#1}
  \refstepcounter{equation}%
  \label{#2}%
  \addtocounter{equation}{-1}%
  \endgroup
}

\makeatletter\def\toclevel@title{9}\makeatother
\hypersetup{bookmarksdepth=8}
\begin{document}
\pagestyle{plain}

\title{Optomechanical Limit on Intracavity Intensities from Bulk Mirror Vibrations}


\author{L.~Maisenbacher}
\altaffiliation{These authors contributed equally to this work.}
\email[Contact author: ]{lothar.maisenbacher@berkeley.edu}
\author{A.~Singh}
\altaffiliation{These authors contributed equally to this work.}
\author{I.~M.~Pope}
\affiliation{Department of Physics, University of California, Berkeley, Berkeley, CA, USA}
\author{H.~Müller}
\affiliation{Department of Physics, University of California, Berkeley, Berkeley, CA, USA}
\affiliation{Lawrence Berkeley National Laboratory, Berkeley, CA, USA}

\date{\today}

\begin{abstract}
Ultrahigh continuous-wave intensities (\SI{>300}{GW/cm^2}) in high-numerical-aperture optical cavities enable applications from phase-contrast electron microscopy to ultradeep dipole traps for molecules.
However, the intensity can be limited by the optomechanical parametric instability (PI), where mirror vibrations scatter light from one cavity mode into another.
We observe parametric instability in a tabletop Fabry-Pérot cavity, show that the mechanical modes are MHz-frequency bulk acoustic modes inside the mirrors, and measure their $Q$ factor.
By using low-$Q$ mirrors, we achieve \SI{>500}{GW/cm^2} intensities in an open, free-space cavity.

\vspace{\baselineskip}
\noindent Published in Phys.\ Rev.\ Lett.\ \textbf{137}, 203603 (2026), DOI: \href{https://doi.org/10.1103/y5kw-s33z}{10.1103/y5kw-s33z}
\end{abstract}


\maketitle
\thispagestyle{plain}

\phantomeq{S1a}{eq: D0 dot}
\phantomeq{S1b}{eq: D1 dot}
\phantomeq{S1c}{eq: X dot}

\phantomsection\addcontentsline{toc}{section}{Introduction}
\textit{Introduction}---Optical cavities with high intracavity power and numerical aperture ($\mathrm{NA} > 0.01$) are an emerging technology that enables ultrahigh continuous-wave intensities (\SI{>300}{GW/cm^2}) \cite{Turnbaugh2021}.
Pioneered for use as a laser phase plate for electron microscopy~\cite{Mueller2010,Schwartz2017,Schwartz2019,Petrov2026}, such cavities may be used for coherent electron beam manipulation~\cite{Axelrod2020} or as ultradeep (\SI{>1}{K}) optical dipole traps for molecules~\cite{Singh2023}.
Their intensities are enabled by careful management of thermal effects through low-expansion mirror substrates and low-absorption, low-defect mirror coatings \cite{Axelrod2026,Turnbaugh2021,Schwartz2019}.
As we show here, the radiation pressure acting on the mirrors can likewise impose an optomechanical limit on the intracavity intensity through the parametric instability~(PI) \cite{Braginsky2001}.
Three-mode PI occurs when a nascent excitation of a mirror's mechanical mode, possibly generated by thermal fluctuations, is amplified by light circulating in the driven mode of the optical cavity \cite{Evans2010}.
If the mechanical mode is resonant with the frequency difference between the driven mode and a higher-order optical mode, and overlaps spatially with both, light is Stokes-scattered from the driven mode into the higher-order mode by mirror vibrations.
The resulting beating between the two optical modes modulates the radiation pressure, which further drives the mechanical mode, closing the feedback loop.
Above a threshold power, the loop's parametric gain exceeds unity, and energy is transferred into the higher-order mode and the mirror vibration, ultimately clamping the driven mode's power at the threshold.
In contrast, two-mode PI scatters light into a frequency sideband of the same optical mode through a mechanical mode with a frequency within the cavity linewidth.

Three-mode PI has been observed in purpose-built cavities incorporating intracavity oscillators \cite{Chen2015, Kharel2019} and in the km-long arm cavities and kg-scale mirrors of the LIGO~gravitational-wave detectors~\cite{Evans2015}.
In the latter, PI is a major limitation on achievable power unless carefully suppressed~\cite{Blair2017,Biscans2019,Hardwick2020}.
Additionally, two-mode PI has been observed in a variety of micromechanical systems \cite{Aspelmeyer2014} and in a 1-m-long version of the LIGO~cavity~\cite{Corbitt2006}.
Here, we report on the first observation of three-mode PI in a tabletop Fabry-Pérot optical cavity, show that it limits the achievable intensity, and demonstrate a suppression scheme.

\begin{figure}[t]
    \includegraphics{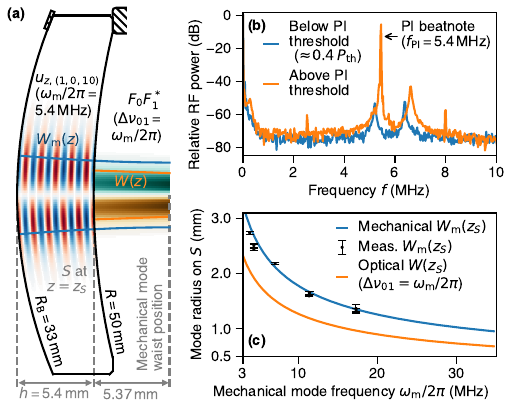}
    \caption{\label{fig:MirrorModeFigure}
    (a)~Cross section of cavity mirror, held by three flexure-mounted pads (hatched, one shown).
    Displacement $u_z$ of a mechanical mode (with indices $(m, n, p)=(1,0,10)$ and frequency $\omega_\m/(2\pi) = \SI{5.4}{MHz}$) shown in blue and red for positive and negative values, respectively.
    Beatnote amplitude $F_0F_1^*$ of the driven and higher-order optical modes for transverse mode spacing $\TMS=\omega_\m/(2\pi)$ shown in green and orange for positive and negative values, respectively.
    (b)~Radio frequency power spectrum of cavity transmission on apertured photodetector.
    A weak laser sideband at frequency $f$ was applied to show the transverse mode spectrum of the (astigmatic) optical cavity.
    Data from ULE~mirror set~2; \SI{10}{kHz}~resolution bandwidth.
    (c)~Calculated (blue line) and measured (black circles) mechanical mode radii $\wmech(\zs)$, and calculated optical mode radii $\wzsopt$ (orange line) at mirror front surface $S$.
    Error bars: 1$\sigma$ uncertainty.
    }
\end{figure}

Our cavity is symmetric and near-concentric, $L \approx \SI{100}{mm}$ in length, and consists of two $\RS = \SI{50}{mm}$ radius of curvature mirrors made from ultra-low-expansion glass~(ULE)~\footnote{Corning ULE 7972, \SI{0.7}{\percent\per\cm} absorption at \SI{1064}{nm}.} (\cref{fig:MirrorModeFigure}(a)).
It operates in vacuum at a wavelength $\lambda = \SI{1064}{nm}$ with circulating powers up to \SI{\approx100}{kW}.
To allow high-NA operation, the mirrors were designed to minimize spherical aberrations of the transmitted beam while forming converging lenses.
However, such a design tends to form stable acoustic cavities inside the mirrors (Supplemental Material~\cite{SM}\nocite{Yoshida1982,Sudhir2016PhDThesis,sdeint,Feldheim1940}), as is the case here.
Their bulk acoustic modes can then interact with the optical modes at \si{MHz} frequencies, which matches the typical mode spacing of tabletop cavities, leading to PI.
This contrasts with the \unit{kHz} surface acoustic modes causing PI in gravitational-wave detectors~\cite{Evans2015}, which do not depend on the bulk mirror geometry.

\phantomsection\addcontentsline{toc}{section}{Theory}
\textit{Theory}---We assume the $\TEMfundamental$ optical mode ($q$: longitudinal index), containing circulating power $\Pcirc$, as the driven mode and the $\TEMHOM$ optical mode as the higher-order mode, where $\mathrm{TEM}_{mn,q}$ are Hermite-Gaussian modes ($m,n$: transverse indices).
The PI~parametric gain for a mechanical mode with index $j$ is $\Pcirc/\PthreshIndex$~\cite{Evans2015}, where $\PthreshIndex$ is a power threshold.
The mechanical mode's exponential decay time is
\begin{equation}\label{eq:DressedDecayTime}
    \tau_\mIndex = 1/[\gamma_\mIndex(1-\Pcirc/\PthreshIndex)],
\end{equation}
with the mechanical amplitude decay constant $\gamma_\mIndex$.
When $\Pcirc$ exceeds $\PthreshIndex$, PI occurs ($\tau_\mIndex<0$), and the mechanical mode amplitude grows exponentially.
$\PthreshIndex$ is given by~\cite{Braginsky2001,Evans2010,NoteBraginsky2001}
\begin{equation}\label{eq: def P threshold}
\PthreshIndex=\frac{c\,\omega_\mIndex^2 L M_j}{4Q_1Q_\mIndex|B_j|^2}\left(1+\frac{\Deltaomega^2}{(\gamma_1+\gamma_\mIndex)^2}\right),
\end{equation}
where $\omega_\mIndex$ is the mechanical mode's angular frequency and $\Deltaomega = \omega_0-\omega_1-\omega_\mIndex$ the detuning ($\omega_0$, $\omega_1$, angular frequencies of $\TEMfundamental$ and $\TEMHOM$ modes; $c$, speed of light).
$Q_1=\omega_1/(2\gamma_1)=\omega_1L\finesse/(\pi c)$ and $Q_\mIndex = \omega_\mIndex/(2\gamma_\mIndex)$ are the $Q$ factors of the $\TEMHOM$ mode and the mechanical mode ($\gamma_1$, $\finesse$, amplitude decay constant and finesse of the $\TEMHOM$ mode).
The effective mass of the mechanical mode is $M_j = \int_V\rho(\bm{r}) |\bm{u}_j(\bm{r})|^2d\bm{r}$, where $\bm{u}_j$ is the mode displacement, $\rho$ the mirror density, and $V$ the mirror volume.
The overlap of the mechanical and optical modes over the mirror surface $S$ is
\begin{equation}\label{eq: def B}
    B_j = \int_S u_{z,j}(\bm r_\perp)F_0(\bm r_\perp)F_1^*(\bm r_\perp)d\bm r_\perp,
\end{equation}
where $u_{z,j}$ is the $z$ component (along cavity axis) of $\bm{u}_j$.
The mode functions $F_0$ and $F_1$ of the $\TEMfundamental$ and $\TEMHOM$ modes are normalized to $\int|F(\bm r_\perp)|^2d\bm r_\perp = 1$.

The geometry of our near-concentric cavity sets its transverse mode spacing,
\begin{equation}\label{eq: def TMS}
    \TMS = \frac{\omega_0-\omega_1}{2\pi} = \frac{c\cos^{-1}\left(\frac{L-\RS}{\RS}\right)}{2\pi L}\approx \frac{c\sqrt{d}}{2\sqrt{2}\pi \RS^{3/2}},
\end{equation}
where $d=2\RS-L$ is the distance to concentricity, which we tune between \num{0.5} and \SI{120}{\micro\meter}.
The mode waist~$\wzeroopt$ and numerical aperture $\NA$ are related to $d$ by $\wzeroopt = \sqrt{\lambda L/2\pi} (d/L)^{\frac{1}{4}}$ and $\NA =\sqrt{2\lambda/\pi L} (L/d)^{\frac{1}{4}}$.

The front surfaces of our mirrors typically show a small astigmatism ($\Delta R \sim \SI{1}{\micro\meter}$), which splits the otherwise degenerate $\TEMHOM$ and $\TEMOHOM$ mode frequencies by $\gtrsim\!\gamma_1/(2\pi)$ (\cref{fig:MirrorModeFigure}(b)).
PI only involves one of these modes at a time (Supplemental Material~\cite{SM}), which we refer to as $\TEMHOM$.

Calculating the power threshold (\cref{eq: def P threshold}) requires knowledge of the relevant mechanical modes.
As detailed in \cref{appendix:BulkAcousticWaveResonators}, these modes can be calculated analytically using the same paraxial formalism that describes optical cavity modes, applied to longitudinally polarized bulk acoustic waves resonating between the two surfaces of our mirrors.
Here, each mirror (\cref{fig:MirrorModeFigure}(a)) behaves as a near-planar acoustic cavity with a nominal free spectral range $\FSRmech = \SI{530(10)}{kHz}$, transverse mode spacing $\TMSmech = \SI{46(1)}{kHz}$, and mode frequencies $\omega_\mIndex/(2\pi) = p\,\FSRmech+(m+n+1)\,\TMSmech$ ($m, n$: transverse indices; $p$: longitudinal index).
The mirror geometry fixes the mechanical mode waist position and Rayleigh range.
Consequently, the mechanical mode radius $\wmechIndex(z)$ scales as $1/\sqrt{\omega_\mIndex}$.
Likewise, the optical mode radius $\wzsopt \approx \lambda L/(2\pi\wzeroopt)$ at the mirror front surface scales as $1/\sqrt{\TMS}$.
For (nearly) resonant modes, i.e., $\omega_\mIndex/(2\pi) \approx \TMS$, all the mode profile dimensions in \cref{eq: def B} hence scale as $1/\sqrt{\omega_\mIndex}$ (\cref{fig:MirrorModeFigure}(c)).
If the optical and mechanical modes are centered and aligned (sharing origin and principal axes orientation, respectively), $|B_j^2|/M_j$ scales as $\omega_\mIndex$ with a mode-specific prefactor.
In this case, only odd $m$ and even $n$ modes can overlap with the beating optical modes, with the strongest overlap for $(m=1, n=0)$ modes.
Importantly, the achievable intensity $I \propto \Pthresh/\wzeroopt^2$ is then independent of $\omega_\mIndex$, i.e., the intensity limit is independent of NA.
The selection rule vanishes for imperfect centering or alignment, but the intensity limit remains approximately NA-independent.

\begin{figure}[b]
    \includegraphics{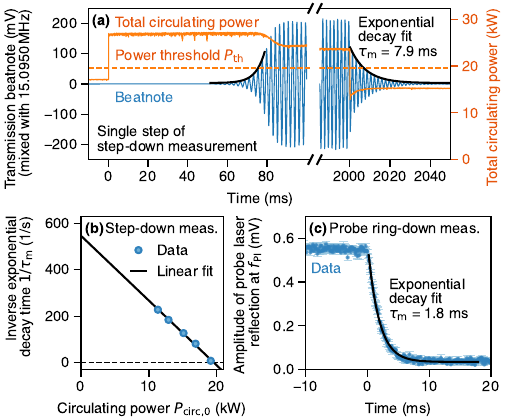}
    \caption{\label{fig:Signals}
    (a)~Single step of a step-down measurement, in which the total circulating power (power $\Pcirc$ in the $\TEMfundamental$ mode plus power in $\TEMHOM$ mode, orange line) is increased above and reduced below the power threshold $\Pthresh$ (dashed orange line).
    Cavity transmission shows a beatnote (blue line; mixed down from \SI{15.1}{MHz} for visibility) with exponentially rising and falling envelope.
    Fitting the falling envelope (black line) gives the mechanical mode's exponential decay time $\tau_\m$ at the final circulating power $\Pcirc$.
    (b)~Linear fit (black line) to the inverse decay times $1/\tau_\m$ (blue circles) at powers $\Pcirc$, giving the extrapolated threshold $\Pthresh$ and $\tau_\m$ at zero power (\cref{eq:DressedDecayTime}).
    (c)~Probe ring-down measurement, in which the total circulating power is reduced from above the threshold $\Pthresh$ to zero (at time \SI{0}{ms}).
    The amplitude (blue circles) of the probe laser's reflection off the cavity at frequency $\PIFreq$, which is proportional to the mirror vibration amplitude, decays exponentially.
    A fit (black line) gives $\tau_\m$.
    Data averaged over seven repetitions. Error bars: 1$\sigma$ uncertainty.
    }
\end{figure}

\begin{figure*}[t!]
    \includegraphics{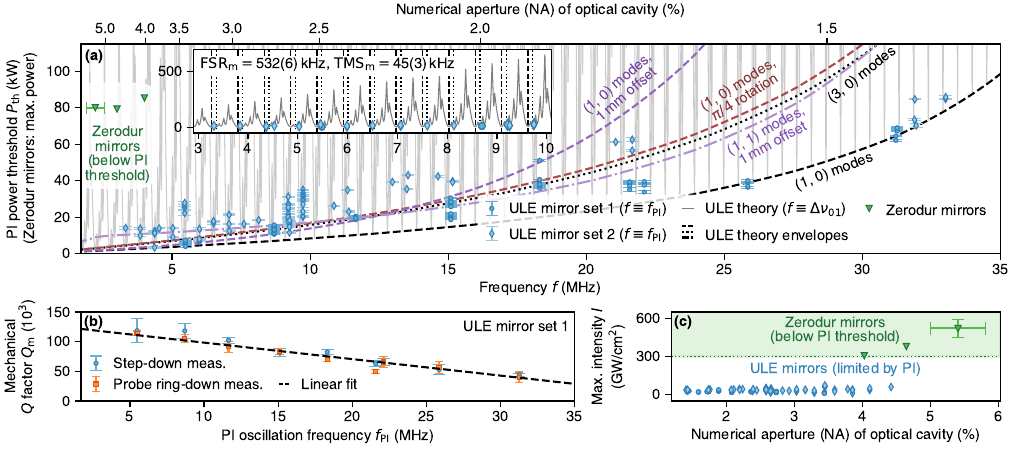}
    \caption{
    \label{fig:PowerThresholdAndQFactor}
    (a)~Measured and calculated PI~power thresholds $\Pthresh$ at frequency $f$ or optical cavity NA ($f = \TMS$).
    Blue circles and diamonds, experimental data ($f$, observed PI~oscillation frequency $\PIFreq$) for two sets of ULE~cavity mirrors;
    gray line, calculated minimum threshold over all mechanical modes, $\min_j \PthreshIndex$, for $f \equiv \TMS$ (optical modes centered and aligned with the mechanical modes);
    dashed and dotted lines, envelopes of the on-resonance ($\TMS = \omega_\mIndex/(2\pi)$) thresholds for the labeled mechanical modes and for radial offsets and rotations of the optical modes; black dashed line, lower envelope of the gray line, which agrees with the lowest measured thresholds;
    green triangles, maximum circulating power with Zerodur mirrors (PI sporadically observed at higher powers but not characterized).
    Finesse set to $\finesse =\num{28000}$; mirror set~2 data ($\finesse = \num{32000}$) scaled accordingly.
    Inset: $\min_j \PthreshIndex$ with experimental $\Pthresh$; dashed and dotted lines mark $(1, 0)$ and $(3, 0)$ mechanical mode frequencies.
    Error bars: 1$\sigma$ statistical uncertainty; \SI{3}{\percent} power calibration uncertainty not included.
    (b)~Mechanical $Q$ factors $Q_\mIndex$ from step-down (blue circles) and probe ring-down (orange squares) measurements.
    Error bars: weighted standard deviation (step-down) and combined statistical and systematic 1$\sigma$ uncertainty (probe ring-down).
    (c)~Maximum intracavity intensity, inferred from circulating power and NA.
    }
\end{figure*}

\phantomsection\addcontentsline{toc}{section}{Experimental scheme}
\textit{Experimental scheme}---During PI, the optical cavity oscillates in both the $\TEMfundamental$ and $\TEMHOM$ modes.
We directly observed their beatnote in the rf power spectrum of the cavity transmission (\cref{fig:MirrorModeFigure}(b); an aperture at the photodetector ensures spatial overlap between the modes).
The beatnote frequency defines the PI~frequency $\PIFreq$.
Because the inverse decay time of the mechanical modes ($\gamma_\mIndex < \SI[per-mode=power]{3000}{\per\second}$) is much smaller than that of the $\TEMHOM$ mode ($\gamma_1 \geq \SI[per-mode=power]{1.5e5}{\per\second}$), $\PIFreq$ equals the oscillating mechanical mode frequency, $\omega_\mIndex/(2\pi)$ (Supplemental Material~\cite{SM}).
Using an acousto-optic modulator (\SI{<2}{\micro\second} rise time), we varied the circulating power $\Pcirc$ between above (transiently) and below the power threshold $\PthreshIndex$ and observed the exponential rise and decay of the beatnote (\cref{fig:Signals}(a)).
By repeating this for multiple values of $\Pcirc < \PthreshIndex$ and using \cref{eq:DressedDecayTime}, we determined $\PthreshIndex$ and the mechanical $Q$ factor $Q_\mIndex$ (``step-down measurement,'' \cref{fig:Signals}(b)).
We selected different mechanical modes by varying the optical transverse mode spacing $\TMS$ (by adjusting the distance to concentricity $d$), and then linearly ramping up the circulating power until we observed PI.
However, because of thermal expansion of the mirrors, we could not independently control the detuning $\Deltaomega = 2\pi\TMS - \omega_\mIndex$, and instead observed that PI set in close to resonance with a mechanical mode ($\TMS \approx \PIFreq$, i.e., $\Deltaomega \approx 0$), where the threshold is lowest.
This thermal self-stabilization of the PI~threshold is discussed in detail in \cref{appendix:ThermalEffects}.
The thermal expansion is visible as a shift of the higher-order modes with circulating power in \cref{fig:MirrorModeFigure}(b).

We also directly observed the mirror vibrations using a probe laser (\num{852} or \SI{948}{nm} wavelength) coupled into the optical cavity.
Because of the low mirror reflectance at these wavelengths (\SIrange{20}{32}{\percent}) and the near-degenerate mode structure, the cavity's spatial selectivity is low, allowing us to probe different positions $(x,y)$ on the mirror front surface.
We stabilized the probe laser to the side of the cavity fringe, and measured the local cavity length change at $\PIFreq$ from the modulation of the reflected light.
Since only one mirror vibrates at a time (Supplemental Material~\cite{SM}), this gives the vibration amplitude at $(x,y)$.
By blocking the 1064-nm light with the acousto-optic modulator, we independently inferred $Q_\mIndex$ from the exponential decay of the vibration amplitude (``probe ring-down measurement,'' \cref{fig:Signals}(c)).

\phantomsection\addcontentsline{toc}{section}{Observation of parametric instability}
\textit{Observation of parametric instability}---Using this approach, we measured the power threshold $\Pthresh$ and mechanical $Q$ factor $Q_\m$ of different mechanical modes.
\Cref{fig:PowerThresholdAndQFactor}(a) shows $\Pthresh$ for two ULE~cavity mirror sets differing primarily in coating absorption (\num{0.1} and \SI{0.7}{ppm} for sets 1 and 2; blue circles and diamonds).
Overall, $\Pthresh$ ranges from \SI{3}{kW} at $\PIFreq \approx \SI{4}{MHz}$ (\SI{4}{\percent}~NA) to \SI{86}{kW} at $\PIFreq \approx \SI{33}{MHz}$ (\SI{1.4}{\percent}~NA).
From the spacing of $\PIFreq$, we infer $\FSRmech = \SI{532(6)}{kHz}$ and $\TMSmech = \SI{45(3)}{kHz}$, in good agreement with the values expected from the mirror geometry (inset).

\Cref{fig:PowerThresholdAndQFactor}(b) shows the corresponding values of $Q_\m$ determined from the step-down (blue circles) and the 948-nm probe ring-down (orange squares) measurements, which agree well.
We find that $Q_\m$ decreases approximately linearly with $\PIFreq$ from \num{5.5} to \SI{31}{MHz}, with a best fit of $Q_\m(\PIFreq) = 1.26(5)\times10^5 -2.8(2)\times10^{-3} \PIFreq/\si{Hz}$ (dashed line).
To our knowledge, this is the first measurement of the $Q$ factor of ULE at \si{MHz}~frequencies.
A previous measurement at \si{kHz}~frequencies~\cite{Numata2004} found a frequency-independent $Q$ factor half the value we extrapolate to this range.
The observed frequency dependence may arise from a combination of structural damping and Rayleigh scattering off material defects~\cite{Saulson1990,MasonThurston1971}, though a detailed characterization is beyond the scope of this work.
Our measured $Q_\m$ is approximately tenfold smaller than that of fused silica at \si{MHz}~frequencies \cite{Startin1998,Numata2004}.
Only data from ULE~mirror set~1 are used to determine $Q_\m$, as set~2's higher absorption and resulting thermal expansion introduced substantial nonlinearities when extrapolating $\tau_\m$ in the step-down measurement.

Using $Q_\m(\PIFreq)$, $\FSRmech$, and $\TMSmech$, we calculate the minimum power threshold, $\min_j \PthreshIndex$, over all mechanical modes (Supplemental Material~\cite{SM}), finding that the contributing modes have $m+n \leq 10$.
The result versus optical transverse mode spacing $\TMS$ is shown in \cref{fig:PowerThresholdAndQFactor}(a) (gray solid line), assuming the optical modes are centered and aligned with the mechanical modes.
As expected from the thermal expansion of the mirrors (\cref{appendix:ThermalEffects}), the experimental values are at resonance with low-$(m, n)$ modes, near the lower limit of the calculated $\PthreshIndex$ (inset).
The lower envelope of $\min_j \PthreshIndex$ occurs at resonance with $(1, 0)$ mechanical modes (dashed black line), and agrees well with the lower envelope of the measured $\Pthresh$ values.
We attribute the remaining variation in $\Pthresh$ to alignment imperfections as well as the excitation of mechanical modes other than the $(1, 0)$ mode (other lines), as indicated by oscillation frequencies spaced by $\TMSmech$ and its multiples (inset).

\begin{figure}[t]
    \includegraphics{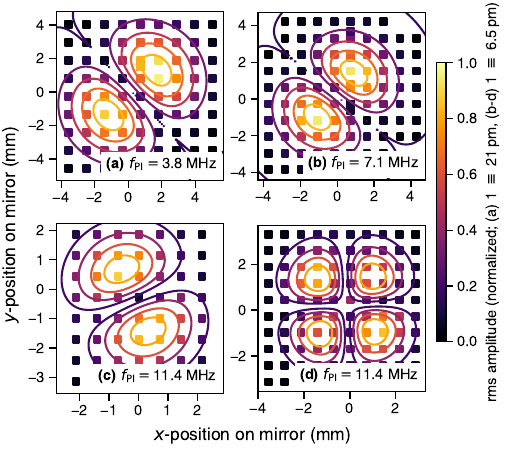}
    \caption{
    \label{fig:ModeScans}
    Vibration amplitude profiles of mechanical modes in the cavity mirrors excited by PI, measured with the 852-nm probe laser.
    Contours show fitted Hermite-Gaussian modes with indices (a)--(c) $(m = 1, n = 0)$ and (d) $(1, 1)$.
    Data from ULE~mirror set~2.
    }
\end{figure}

\phantomsection\addcontentsline{toc}{section}{Measurement of mechanical mode profiles}
\textit{Measurement of mechanical mode profiles}---The vibration amplitude profiles of four of the mechanical modes driven by PI are shown in \cref{fig:ModeScans}, recorded by scanning the 852-nm probe laser (\SI{0.5}{mm} beam radius) across the cavity mirrors.
These profiles are well described by Hermite-Gaussian modes with indices $(1, 0)$ (a)--(c) and $(1, 1)$ (d), and their amplitudes match the order of magnitude expected from our theory.
This agreement holds for four additional measurements not shown here.
The principal axes of the optical modes (measured in transmission on a camera) are approximately aligned with those of the mechanical modes.
The fitted mechanical mode radii $\wmech(\zs)$ (\cref{fig:MirrorModeFigure}(c)) agree within \SI{4}{\percent} with those expected from the mirror geometry.
The profile measurement might be biased toward lower-$(m,n)$ modes compared to our other measurements, as the longer measurement time selects for modes where PI can be reliably reproduced.

\phantomsection\addcontentsline{toc}{section}{Suppression of parametric instability}
\textit{Suppression of parametric instability}---PI can be suppressed without changing the properties of the optical cavity through several approaches.
These include controlling the detuning $\Deltaomega$ to operate far from a mechanical resonance~\cite{Evans2015,Hardwick2020}, decreasing the mechanical $Q$ factor through damping~\cite{Blair2017,Biscans2019} or low-$Q$ mirror substrates, or modifying the mirror geometry to render the acoustic cavity unstable.
The latter, combined with a high $\TMS \approx \SI{100}{MHz}$, likely suppressed PI in the (20-\si{mm}-long, largely enclosed) laser-phase-plate cavities~\cite{Turnbaugh2021,Axelrod2024Thesis}, whose short length permits a low-aberration ULE~mirror geometry that does not form a stable acoustic cavity (Supplemental Material~\cite{SM}).
In an otherwise identical test cavity whose mirrors formed stable acoustic cavities, we believe we observed PI at a threshold consistent with our results above.

Here, we reduce the mechanical $Q$ factor, as control of the detuning is limited by thermal expansion, and the PI~threshold is challenging to predict for unstable geometries.
However, the localized nature of our mechanical modes makes damping challenging without affecting the optical modes, and attempts using an indium damper failed to provide sufficient damping and distorted the mirrors.
Instead, we replaced our ULE~mirrors with mirrors made from ultra-low-expansion glass-ceramic Zerodur~\footnote{Schott Zerodur Expansion Class 0 Special.} but with an otherwise identical geometry, superpolishing, and coating design.
Zerodur has a 20-fold lower $Q$ factor than ULE at \si{kHz}~frequencies~\cite{Numata2004}.
With these mirrors ($\finesse = \num{26000}$), we achieved a maximum circulating power of \SI{85}{kW} at $\TMS = \SI{4.0}{MHz}$ (\SI{4.0}{\percent} NA, \SI{310}{GW/cm^2} intensity; green triangles in Figs.~\ref{fig:PowerThresholdAndQFactor}(a) and \ref{fig:PowerThresholdAndQFactor}(c)).
The circulating power was limited by thermal lensing from Zerodur's large absorption (\SI{3.4}{\percent\per\cm} at \SI{1064}{nm}), which reduced coupling efficiency at higher input power.
Higher intensities were realized at higher NAs, reaching \SI{380}{GW/cm^2} at \SI{4.6}{\percent} NA before the onset of strong aberrations.
Beyond $\text{NA} \approx \SI{5}{\percent}$, the optical cavity mode was highly aberrated and the NA could no longer be accurately determined from the transverse mode spectrum; instead, we imaged the mode (correcting for thermal lensing by varying the circulating power) to find the peak intensity $I$ and corresponding effective NA, $\sqrt{I \lambda^2/(8\pi\Pcirc)}$.
We achieved hour-long operation with up to \SI{520(70)}{GW/cm^2} at an effective NA of \SI{5.4(0.4)}{\percent}, comparable to the highest intensities demonstrated in~\cite{Turnbaugh2021,Axelrod2024Thesis}, despite our fivefold smaller distance to concentricity.
We sporadically observed PI at $\Pthresh\approx\SI{80}{kW}$ and $\TMS \approx\SI{0.9}{MHz}$, but were unable to measure the $Q$ factor because the onset of PI was not reliably reproducible.
However, the threshold suggests a $Q$ factor of $\SIrange{e3}{e4}{}$, in line with the value \num{3.1e3} reported in \cite{Numata2004}.
Even higher intensities are likely achievable with ULE~mirrors whose geometry does not form a stable acoustic cavity, although care needs to be taken to minimize optical aberrations (Supplemental Material~\cite{SM}).

\bigskip
\phantomsection\addcontentsline{toc}{section}{Conclusions}
\textit{Conclusions}---We showed that bulk mirror vibrations at \si{MHz}~frequencies can limit the power and intensity of high-NA Fabry-Pérot cavities through PI.
In particular, cavity mirrors optimized for low optical aberrations tend to form stable acoustic cavities that have localized mechanical modes leading to a relatively low power threshold for PI, exacerbated by the high $Q$ factors of optical glasses.
We measured the power thresholds, the $Q$ factors, and the profiles of the mechanical modes in our system, finding good agreement with the corresponding theory.
These results place mirror geometry or substrate limitations on any high-NA, tabletop Fabry-Pérot cavity.
The localized nature of the modes lets PI probe intrinsic mechanical $Q$ factors at \si{MHz}~frequencies, which we used to determine the $Q$ factor of ULE.
By externally driving the higher-order optical mode, this technique could also be used without reaching the PI~threshold.
Finally, we demonstrated that low-$Q$ mirrors (made from Zerodur) suppress PI, allowing us to reach intensities exceeding \SI{500}{GW/cm^2} in an open, free-space cavity.
Notably, such intensities enable ultradeep, buffer-gas-loaded dipole traps for molecules, including for low-polarizability species like atomic and molecular hydrogen~\cite{Singh2023}.

\bigskip
\begin{acknowledgments}
\textit{Acknowledgments}---The authors thank E.~S.~Cooper for insightful theory discussions, J.~J.~Axelrod for help with the setup of the test cavity, S.~Straßer and P.~Stromberger for help with developing a mirror inspection microscope, P.~N.~Petrov and J.~T.~Zhang for helpful discussions, and R.~Patterson for technical advice on optical coatings.
This work was supported by the Brown Science Foundation through the Brown Investigator Award (Award No.~1167), the Gordon and Betty Moore Foundation (Grant No.~9366), the Chan Zuckerberg Initiative (Awards No.~2021-234606 and No.~2025-367757), the Heising-Simons Foundation (Grant No.~2023-4467), and the U.S.~Department of Energy, Office of Science, National Quantum Information Science Research Centers, Quantum Systems Accelerator (Award No.~DE-SCL0000121).
L.~M. acknowledges a Feodor Lynen Fellowship from the Alexander von Humboldt Foundation.
A.~S. acknowledges support from the UC~Berkeley Physics Department's Graduate Student Support Fund.
\end{acknowledgments}

\bigskip
\phantomsection\addcontentsline{toc}{section}{Data availability}
\textit{Data availability}---The data shown in the figures, including those of Supplemental Material~\cite{SM}, are available from Ref.~\cite{DataPI2026}. All other data, and the code used for the analysis and the simulations, are available from the authors upon request.

%

\appendix
\onecolumngrid
\begin{center}
\textbf{End Matter}
\end{center}
\twocolumngrid

\appendixheading{appendix:BulkAcousticWaveResonators}{Paraxial theory of bulk acoustic modes in cavity mirrors}The acoustic modes relevant to our observed parametric instability can be calculated explicitly by means of a paraxial approximation to the vibrational wave equation.
Here, we derive this approximation from first principles.
This approach is similar to that used in~\cite{Kharel2018, Renninger2018}, but is simplified by the fact that our ULE mirrors are isotropic.

Mechanical modes can often be well described as surface waves, mostly transverse waves in the bulk, or mostly longitudinal waves in the bulk.
From calculations not shown here, we find that the mode overlaps of the transverse or surface waves at our observed PI frequencies are too small to explain the observed PI in this Letter.
Longitudinal bulk acoustic waves describe the relevant modes observed here.
The displacement $\bm u$ of longitudinal waves can be written in terms of a scalar potential $\Phi$ as $\bm u = \nabla \Phi$~\cite{LayWallace1995}, where $\Phi$ satisfies the wave equation
\begin{align}\label{eq: longitudinal wave equation}
    c_\mathrm{L}^2\nabla^2\Phi-\ddot\Phi = 0
\end{align}
and appropriate boundary conditions.
Here, $c_\mathrm{L}$ is the longitudinal wave velocity (\SI{5740}{m/s} for ULE).

The modes relevant to the observed PI must overlap well with the optical $\TEMfundamental$ and $\TEMHOM$ modes (which have transverse length scales of \si{mm}), but at several MHz frequencies the wavelengths of the longitudinal modes will be $\lesssim\!\SI{500}{\micro\meter}$.
This justifies a paraxial approximation for the mechanical mode functions \cite{Cook1976,Newberry1989}.
Formally, we assume a set of solutions to \cref{eq: longitudinal wave equation} of the form $\Phi_j = U_j(x, y, z)e^{i(k_\mIndex z-\omega_\mIndex t)}$, where $U_j$ is a slowly varying envelope function, and $z$ points along the mirror's axis of symmetry.
Following the treatment of~\cite{Lax1975}, we find the acoustic paraxial wave equation
\begin{equation}\label{eq: paraxial wave equation}
    \nabla_\perp^2 U_j + 2ik_\mIndex \partial_z U_j
 = 0\end{equation}
with the constraint $k_\mIndex=\pm\omega_\mIndex/c_\mathrm{L}$.

The bulk acoustic modes must satisfy both \cref{eq: paraxial wave equation} and the boundary conditions of the mirror. The transverse boundary conditions, defined along the edges of the mirror and the mount, are complicated, but can be satisfied straightforwardly by Hermite-Gaussian (HG) modes, which are solutions to \cref{eq: paraxial wave equation} with negligible extent at the edges of the mirror.
Non-Gaussian, delocalized acoustic modes of the mirror also exist.
However, only the HG~modes are responsible for the observed PI, as they have higher overlap $|B_j^2|/M_j$ with the optical modes, and likely have a higher $Q_\mIndex$ since their amplitude is negligible at the mirror's mounting points (hatched elements in \cref{fig:MirrorModeFigure}(a); contacting mirror at radius $r > \SI{11.6}{mm}$).
Because of this localization, we estimate that mounting losses place an upper limit of \num{\sim2e8} on $Q_\mIndex$ for the modes observed here.

Meanwhile, the mirror's front and back surfaces have stress-free boundary conditions,  $\sigma_{xz} = \sigma_{yz} = \sigma_{zz} = 0$ \cite{Rasband1975}.
Given a mechanical mode's displacement, we can compute the stress tensor components of HG modes on these surfaces using the stress-strain relation for a linear, isotropic, elastic solid. Then, within the paraxial approximation, we can expand these components in orders of the paraxial parameter (which, for Gaussian modes, is the mode waist divided by twice its Rayleigh length \cite{Lax1975}).
To leading order in the paraxial parameter, we find $\left(\sigma_{xz}, \sigma_{yz}, \sigma_{zz}\right)\approx\left(0, 0, -k_\mIndex^2 c_\mathrm{L}^2 \rho \Phi_j \right)$, where $\rho$ is the mirror density.
The off-axis, normal stress components on the spherical mirror surfaces will approximately take the same form, provided the wavefront curvature of the HG~modes matches the curvature of the surfaces.
This gives the approximate boundary condition $\Phi_j=0$ (i.e., $u_{z,j}$ has an antinode) on the mirror's front and back surfaces.
The boundary conditions for $\Phi_j$ are identical to those of an optical cavity, and the mechanical HG~mode profiles therefore have the same form as the optical $\mathrm{TEM}$ mode profiles, albeit with different beam parameters.
This fact allows us to write down the modes and their properties explicitly, as is done in Supplemental Material~\cite{SM}.

\appendixheading{appendix:ThermalEffects}{Thermal effects}With the high circulating powers used in our experiments, even small amounts of absorption in the mirror coating can lead to significant thermal effects.
In particular, the thermal expansion of the mirrors increases the distance to concentricity $d=2R-L$ and hence changes the optical mode spacing (\cref{eq: def TMS}) \cite{Axelrod2026}.
Despite using mirrors made from ULE, we find that the optical mode spacing can shift on the order of several optical linewidths as the circulating power is increased to the PI~threshold.
This can be approximately modeled by keeping the differential equations for the optical and mechanical fields, \cref{eq: D0 dot,eq: D1 dot,eq: X dot} of Supplemental Material~\cite{SM}, and allowing $\Delta\omega(t)$ to be a dynamical variable.
$\Delta\omega(t)$ will exponentially approach, with a rate $\alpha$, a set value which depends on the circulating power,
\begin{equation}\label{eq: Delta omega dot}
    \dot{(\Delta\omega)}  = \alpha(\DeltaomegaInit+\beta\Pcirc-\Delta\omega(t)),
\end{equation}
where $\DeltaomegaInit$ is the initial value of $\Delta\omega(t)$ at zero circulating power $\Pcirc$.
Here, $\beta$ is a coefficient representing linear changes in the transverse mode spacing introduced by thermal expansion of the mirrors \cite{Axelrod2026}.
Power in the $\TEMHOM$ is ignored, as it is usually significantly lower than $\Pcirc$, and is spread over a larger area on the mirror.
From empirical observations, the rate constant $\alpha$ is of order \SI{1}{s^{-1}}. Since the timescale of the dynamics of the optical and mechanical fields is significantly faster than $1/\alpha$, at any given $\Delta\omega(t)$ a power threshold is defined by
\begin{equation}
    \Pthresh(t) = \Pthreshzero\left(1+\frac{\Delta\omega^2(t)}{(\gamma_1+\gamma_\m)^2}\right),
\end{equation}
where $\Pthreshzero = c\omega_\m^2LM/(4Q_1Q_\m|B|^2)$.

Now suppose the $\TEMfundamental$ mode is driven well above its power threshold at some $\Delta\omega(t)$. $\Pcirc$ is pinned at $\Pthresh(t)$, which we can insert into \cref{eq: Delta omega dot} to find an equation for the slow-timescale dynamics,
\begin{equation}\label{eq: slow time dynamics delta omega}
    \frac{1}{\alpha}\frac{\mathrm{d}\Delta\omega}{\mathrm{d}t} = \frac{\beta\Pthreshzero\Delta\omega^2(t)}{(\gamma_1+\gamma_\m)^2}-\Delta\omega(t)+(\DeltaomegaInit+\beta\Pthreshzero).
\end{equation}

For a thermally self-stable equilibrium of \cref{eq: slow time dynamics delta omega} to exist, the discriminant $\Delta_{\mathrm{therm}} = 1-4\beta\Pthreshzero(\DeltaomegaInit+\beta \Pthreshzero)/(\gamma_1+\gamma_\m)^2$ must be positive. In this case, the stable equilibrium is
\begin{equation}\label{eq: delta omega stable equilibrium solution}
    \Deltaomegaeq = \frac{1-\sqrt{\Delta_{\mathrm{therm}}}}{2\beta\Pthreshzero/(\gamma_1+\gamma_\m)^2}\leq\frac{(\gamma_1+\gamma_\m)^2}{2\beta\Pthreshzero}.
\end{equation}
It is therefore impossible to drive PI at a detuning above $\Delta\omega=(\gamma_1+\gamma_\m)^2/2\beta\Pthreshzero$.
If the cavity geometry is adjusted so that $\Delta\omega$ exceeds this limit, the cavity circulating power can be increased until the optical mode spacing shifts into near resonance with another mechanical mode, which will then begin to oscillate instead.

We empirically find $\beta$ to have an approximately linear frequency dependence, with $\beta\sim\gamma_1/\Pthreshzero$ at $\PIFreq\sim \SI{5}{MHz}$ and $\beta\sim70\gamma_1/\Pthreshzero$ at $\PIFreq\sim\SI{30}{MHz}$ in ULE~mirror set~2 (\cref{fig:MirrorModeFigure}(b)), and approximately sevenfold lower for mirror set~1 owing to its lower coating absorption.
For most of our experimental range, $\beta$ is comparable to or larger than $\gamma_1/\Pthreshzero$, constraining the detuning at the onset of PI to $\Delta\omega \lesssim\gamma_1/2$ (\cref{eq: delta omega stable equilibrium solution}) and favoring low-threshold mechanical modes at whichever longitudinal index $p$ satisfies this constraint.
For this reason, the experimental values of $\Pthresh$ cluster close to the lower envelope of the calculated thresholds for low $(m,n)$ and hence low-threshold modes (\cref{fig:PowerThresholdAndQFactor}(a)).
Only at the lowest observed PI~frequencies in mirror set~1 do we find that $\beta$ is small enough, in principle, to control $\Delta\omega$ independently of the circulating power.
However, even in this regime we were unable to increase the circulating power significantly above the lower envelope in \cref{fig:PowerThresholdAndQFactor}(a).
Explaining this observation likely requires accounting for the combined influence of all mechanical modes in both mirrors, which is beyond the scope of this work.
We note that we did not attempt to maximize the circulating power, e.g., by finding an optimal power ramping scheme, as thermal drifts would likely preclude stable long-term operation.


\begin{thebibliography}{39}%
\makeatletter
\providecommand \@ifxundefined [1]{%
 \@ifx{#1\undefined}
}%
\providecommand \@ifnum [1]{%
 \ifnum #1\expandafter \@firstoftwo
 \else \expandafter \@secondoftwo
 \fi
}%
\providecommand \@ifx [1]{%
 \ifx #1\expandafter \@firstoftwo
 \else \expandafter \@secondoftwo
 \fi
}%
\providecommand \natexlab [1]{#1}%
\providecommand \enquote  [1]{``#1''}%
\providecommand \bibnamefont  [1]{#1}%
\providecommand \bibfnamefont [1]{#1}%
\providecommand \citenamefont [1]{#1}%
\providecommand \href@noop [0]{\@secondoftwo}%
\providecommand \href [0]{\begingroup \@sanitize@url \@href}%
\providecommand \@href[1]{\@@startlink{#1}\@@href}%
\providecommand \@@href[1]{\endgroup#1\@@endlink}%
\providecommand \@sanitize@url [0]{\catcode `\\12\catcode `\$12\catcode
  `\&12\catcode `\#12\catcode `\^12\catcode `\_12\catcode `\%12\relax}%
\providecommand \@@startlink[1]{}%
\providecommand \@@endlink[0]{}%
\providecommand \url  [0]{\begingroup\@sanitize@url \@url }%
\providecommand \@url [1]{\endgroup\@href {#1}{\urlprefix }}%
\providecommand \urlprefix  [0]{URL }%
\providecommand \Eprint [0]{\href }%
\providecommand \doibase [0]{https://doi.org/}%
\providecommand \selectlanguage [0]{\@gobble}%
\providecommand \bibinfo  [0]{\@secondoftwo}%
\providecommand \bibfield  [0]{\@secondoftwo}%
\providecommand \translation [1]{[#1]}%
\providecommand \BibitemOpen [0]{}%
\providecommand \bibitemStop [0]{}%
\providecommand \bibitemNoStop [0]{.\EOS\space}%
\providecommand \EOS [0]{\spacefactor3000\relax}%
\providecommand \BibitemShut  [1]{\csname bibitem#1\endcsname}%
\let\auto@bib@innerbib\@empty
\bibitem [{\citenamefont {Turnbaugh}\ \emph {et~al.}(2021)\citenamefont
  {Turnbaugh}, \citenamefont {Axelrod}, \citenamefont {Campbell}, \citenamefont
  {Dioquino}, \citenamefont {Petrov}, \citenamefont {Remis}, \citenamefont
  {Schwartz}, \citenamefont {Yu}, \citenamefont {Cheng}, \citenamefont
  {Glaeser},\ and\ \citenamefont {Mueller}}]{Turnbaugh2021}%
  \BibitemOpen
  \bibfield  {author} {\bibinfo {author} {\bibfnamefont {C.}~\bibnamefont
  {Turnbaugh}}, \bibinfo {author} {\bibfnamefont {J.~J.}\ \bibnamefont
  {Axelrod}}, \bibinfo {author} {\bibfnamefont {S.~L.}\ \bibnamefont
  {Campbell}}, \bibinfo {author} {\bibfnamefont {J.~Y.}\ \bibnamefont
  {Dioquino}}, \bibinfo {author} {\bibfnamefont {P.~N.}\ \bibnamefont
  {Petrov}}, \bibinfo {author} {\bibfnamefont {J.}~\bibnamefont {Remis}},
  \bibinfo {author} {\bibfnamefont {O.}~\bibnamefont {Schwartz}}, \bibinfo
  {author} {\bibfnamefont {Z.}~\bibnamefont {Yu}}, \bibinfo {author}
  {\bibfnamefont {Y.}~\bibnamefont {Cheng}}, \bibinfo {author} {\bibfnamefont
  {R.~M.}\ \bibnamefont {Glaeser}},\ and\ \bibinfo {author} {\bibfnamefont
  {H.}~\bibnamefont {Mueller}},\ }\bibfield  {title} {\bibinfo {title}
  {High-power near-concentric {{Fabry}}--{{Perot}} cavity for phase contrast
  electron microscopy},\ }\href {https://doi.org/10.1063/5.0045496} {\bibfield
  {journal} {\bibinfo  {journal} {Rev. Sci. Instrum.}\ }\textbf {\bibinfo
  {volume} {92}},\ \bibinfo {pages} {053005} (\bibinfo {year}
  {2021})}\BibitemShut {NoStop}%
\bibitem [{\citenamefont {M{\"u}ller}\ \emph {et~al.}(2010)\citenamefont
  {M{\"u}ller}, \citenamefont {Jin}, \citenamefont {Danev}, \citenamefont
  {Spence}, \citenamefont {Padmore},\ and\ \citenamefont
  {Glaeser}}]{Mueller2010}%
  \BibitemOpen
  \bibfield  {author} {\bibinfo {author} {\bibfnamefont {H.}~\bibnamefont
  {M{\"u}ller}}, \bibinfo {author} {\bibfnamefont {J.}~\bibnamefont {Jin}},
  \bibinfo {author} {\bibfnamefont {R.}~\bibnamefont {Danev}}, \bibinfo
  {author} {\bibfnamefont {J.}~\bibnamefont {Spence}}, \bibinfo {author}
  {\bibfnamefont {H.}~\bibnamefont {Padmore}},\ and\ \bibinfo {author}
  {\bibfnamefont {R.~M.}\ \bibnamefont {Glaeser}},\ }\bibfield  {title}
  {\bibinfo {title} {Design of an electron microscope phase plate using a
  focused continuous-wave laser},\ }\href
  {https://doi.org/10.1088/1367-2630/12/7/073011} {\bibfield  {journal}
  {\bibinfo  {journal} {New J. Phys.}\ }\textbf {\bibinfo {volume} {12}},\
  \bibinfo {pages} {073011} (\bibinfo {year} {2010})}\BibitemShut {NoStop}%
\bibitem [{\citenamefont {Schwartz}\ \emph {et~al.}(2017)\citenamefont
  {Schwartz}, \citenamefont {Axelrod}, \citenamefont {Tuthill}, \citenamefont
  {Haslinger}, \citenamefont {Ophus}, \citenamefont {Glaeser},\ and\
  \citenamefont {M{\"u}ller}}]{Schwartz2017}%
  \BibitemOpen
  \bibfield  {author} {\bibinfo {author} {\bibfnamefont {O.}~\bibnamefont
  {Schwartz}}, \bibinfo {author} {\bibfnamefont {J.}~\bibnamefont {Axelrod}},
  \bibinfo {author} {\bibfnamefont {D.~R.}\ \bibnamefont {Tuthill}}, \bibinfo
  {author} {\bibfnamefont {P.}~\bibnamefont {Haslinger}}, \bibinfo {author}
  {\bibfnamefont {C.}~\bibnamefont {Ophus}}, \bibinfo {author} {\bibfnamefont
  {R.}~\bibnamefont {Glaeser}},\ and\ \bibinfo {author} {\bibfnamefont
  {H.}~\bibnamefont {M{\"u}ller}},\ }\bibfield  {title} {\bibinfo {title}
  {Near-concentric {{Fabry-P\'erot}} cavity for continuous-wave laser control
  of electron waves},\ }\href {https://doi.org/10.1364/OE.25.014453} {\bibfield
   {journal} {\bibinfo  {journal} {Opt. Express}\ }\textbf {\bibinfo {volume}
  {25}},\ \bibinfo {pages} {14453} (\bibinfo {year} {2017})}\BibitemShut
  {NoStop}%
\bibitem [{\citenamefont {Schwartz}\ \emph {et~al.}(2019)\citenamefont
  {Schwartz}, \citenamefont {Axelrod}, \citenamefont {Campbell}, \citenamefont
  {Turnbaugh}, \citenamefont {Glaeser},\ and\ \citenamefont
  {M{\"u}ller}}]{Schwartz2019}%
  \BibitemOpen
  \bibfield  {author} {\bibinfo {author} {\bibfnamefont {O.}~\bibnamefont
  {Schwartz}}, \bibinfo {author} {\bibfnamefont {J.~J.}\ \bibnamefont
  {Axelrod}}, \bibinfo {author} {\bibfnamefont {S.~L.}\ \bibnamefont
  {Campbell}}, \bibinfo {author} {\bibfnamefont {C.}~\bibnamefont {Turnbaugh}},
  \bibinfo {author} {\bibfnamefont {R.~M.}\ \bibnamefont {Glaeser}},\ and\
  \bibinfo {author} {\bibfnamefont {H.}~\bibnamefont {M{\"u}ller}},\ }\bibfield
   {title} {\bibinfo {title} {Laser phase plate for transmission electron
  microscopy},\ }\href {https://doi.org/10.1038/s41592-019-0552-2} {\bibfield
  {journal} {\bibinfo  {journal} {Nat. Methods}\ }\textbf {\bibinfo {volume}
  {16}},\ \bibinfo {pages} {1016} (\bibinfo {year} {2019})}\BibitemShut
  {NoStop}%
\bibitem [{\citenamefont {Petrov}\ \emph {et~al.}(2026)\citenamefont {Petrov},
  \citenamefont {Zhang}, \citenamefont {Remis}, \citenamefont {Axelrod},
  \citenamefont {Cheng}, \citenamefont {Cooper}, \citenamefont {Hicklin},
  \citenamefont {Sandhaus}, \citenamefont {Schnurr}, \citenamefont {Glaeser},\
  and\ \citenamefont {M{\"u}ller}}]{Petrov2026}%
  \BibitemOpen
  \bibfield  {author} {\bibinfo {author} {\bibfnamefont {P.~N.}\ \bibnamefont
  {Petrov}}, \bibinfo {author} {\bibfnamefont {J.~T.}\ \bibnamefont {Zhang}},
  \bibinfo {author} {\bibfnamefont {J.}~\bibnamefont {Remis}}, \bibinfo
  {author} {\bibfnamefont {J.~J.}\ \bibnamefont {Axelrod}}, \bibinfo {author}
  {\bibfnamefont {H.}~\bibnamefont {Cheng}}, \bibinfo {author} {\bibfnamefont
  {E.~S.}\ \bibnamefont {Cooper}}, \bibinfo {author} {\bibfnamefont {I.~K.}\
  \bibnamefont {Hicklin}}, \bibinfo {author} {\bibfnamefont {S.}~\bibnamefont
  {Sandhaus}}, \bibinfo {author} {\bibfnamefont {C.}~\bibnamefont {Schnurr}},
  \bibinfo {author} {\bibfnamefont {R.~M.}\ \bibnamefont {Glaeser}},\ and\
  \bibinfo {author} {\bibfnamefont {H.}~\bibnamefont {M{\"u}ller}},\ }\bibfield
   {title} {\bibinfo {title} {Laser phase plate improves structure
  determination of small proteins by cryo-{{EM}}},\ }\href
  {https://doi.org/10.1126/science.aeh0665} {\bibfield  {journal} {\bibinfo
  {journal} {Science}\ }\textbf {\bibinfo {volume} {393}},\ \bibinfo {pages}
  {195} (\bibinfo {year} {2026})}\BibitemShut {NoStop}%
\bibitem [{\citenamefont {Axelrod}\ \emph {et~al.}(2020)\citenamefont
  {Axelrod}, \citenamefont {Campbell}, \citenamefont {Schwartz}, \citenamefont
  {Turnbaugh}, \citenamefont {Glaeser},\ and\ \citenamefont
  {M{\"u}ller}}]{Axelrod2020}%
  \BibitemOpen
  \bibfield  {author} {\bibinfo {author} {\bibfnamefont {J.~J.}\ \bibnamefont
  {Axelrod}}, \bibinfo {author} {\bibfnamefont {S.~L.}\ \bibnamefont
  {Campbell}}, \bibinfo {author} {\bibfnamefont {O.}~\bibnamefont {Schwartz}},
  \bibinfo {author} {\bibfnamefont {C.}~\bibnamefont {Turnbaugh}}, \bibinfo
  {author} {\bibfnamefont {R.~M.}\ \bibnamefont {Glaeser}},\ and\ \bibinfo
  {author} {\bibfnamefont {H.}~\bibnamefont {M{\"u}ller}},\ }\bibfield  {title}
  {\bibinfo {title} {Observation of the relativistic reversal of the
  ponderomotive potential},\ }\href
  {https://doi.org/10.1103/PhysRevLett.124.174801} {\bibfield  {journal}
  {\bibinfo  {journal} {Phys. Rev. Lett.}\ }\textbf {\bibinfo {volume} {124}},\
  \bibinfo {pages} {174801} (\bibinfo {year} {2020})}\BibitemShut {NoStop}%
\bibitem [{\citenamefont {Singh}\ \emph {et~al.}(2023)\citenamefont {Singh},
  \citenamefont {Maisenbacher}, \citenamefont {Lin}, \citenamefont {Axelrod},
  \citenamefont {Panda},\ and\ \citenamefont {M{\"u}ller}}]{Singh2023}%
  \BibitemOpen
  \bibfield  {author} {\bibinfo {author} {\bibfnamefont {A.}~\bibnamefont
  {Singh}}, \bibinfo {author} {\bibfnamefont {L.}~\bibnamefont {Maisenbacher}},
  \bibinfo {author} {\bibfnamefont {Z.}~\bibnamefont {Lin}}, \bibinfo {author}
  {\bibfnamefont {J.~J.}\ \bibnamefont {Axelrod}}, \bibinfo {author}
  {\bibfnamefont {C.~D.}\ \bibnamefont {Panda}},\ and\ \bibinfo {author}
  {\bibfnamefont {H.}~\bibnamefont {M{\"u}ller}},\ }\bibfield  {title}
  {\bibinfo {title} {Dynamics of a buffer-gas-loaded, deep optical trap for
  molecules},\ }\href {https://doi.org/10.1103/PhysRevResearch.5.033008}
  {\bibfield  {journal} {\bibinfo  {journal} {Phys. Rev. Res.}\ }\textbf
  {\bibinfo {volume} {5}},\ \bibinfo {pages} {033008} (\bibinfo {year}
  {2023})}\BibitemShut {NoStop}%
\bibitem [{\citenamefont {Axelrod}\ \emph {et~al.}(2026)\citenamefont
  {Axelrod}, \citenamefont {Maisenbacher}, \citenamefont {Singh}, \citenamefont
  {Pope}, \citenamefont {Petrov}, \citenamefont {Zhang},\ and\ \citenamefont
  {M{\"u}ller}}]{Axelrod2026}%
  \BibitemOpen
  \bibfield  {author} {\bibinfo {author} {\bibfnamefont {J.~J.}\ \bibnamefont
  {Axelrod}}, \bibinfo {author} {\bibfnamefont {L.}~\bibnamefont
  {Maisenbacher}}, \bibinfo {author} {\bibfnamefont {A.}~\bibnamefont {Singh}},
  \bibinfo {author} {\bibfnamefont {I.~M.}\ \bibnamefont {Pope}}, \bibinfo
  {author} {\bibfnamefont {P.~N.}\ \bibnamefont {Petrov}}, \bibinfo {author}
  {\bibfnamefont {J.~T.}\ \bibnamefont {Zhang}},\ and\ \bibinfo {author}
  {\bibfnamefont {H.}~\bibnamefont {M{\"u}ller}},\ }\bibfield  {title}
  {\bibinfo {title} {Thermoelastic limit on the focal intensity in
  {{Fabry-P\'erot}} cavities},\ }\href {https://doi.org/10.1103/hbqy-z81p}
  {\bibfield  {journal} {\bibinfo  {journal} {Phys. Rev. A}\ }\textbf {\bibinfo
  {volume} {114}},\ \bibinfo {pages} {043507} (\bibinfo {year}
  {2026})}\BibitemShut {NoStop}%
\bibitem [{\citenamefont {Braginsky}\ \emph {et~al.}(2001)\citenamefont
  {Braginsky}, \citenamefont {Strigin},\ and\ \citenamefont
  {Vyatchanin}}]{Braginsky2001}%
  \BibitemOpen
  \bibfield  {author} {\bibinfo {author} {\bibfnamefont {V.~B.}\ \bibnamefont
  {Braginsky}}, \bibinfo {author} {\bibfnamefont {S.~E.}\ \bibnamefont
  {Strigin}},\ and\ \bibinfo {author} {\bibfnamefont {S.~P.}\ \bibnamefont
  {Vyatchanin}},\ }\bibfield  {title} {\bibinfo {title} {Parametric oscillatory
  instability in {{Fabry}}--{{Perot}} interferometer},\ }\href
  {https://doi.org/10.1016/S0375-9601(01)00510-2} {\bibfield  {journal}
  {\bibinfo  {journal} {Phys. Lett. A}\ }\textbf {\bibinfo {volume} {287}},\
  \bibinfo {pages} {331} (\bibinfo {year} {2001})}\BibitemShut {NoStop}%
\bibitem [{\citenamefont {Evans}\ \emph {et~al.}(2010)\citenamefont {Evans},
  \citenamefont {Barsotti},\ and\ \citenamefont {Fritschel}}]{Evans2010}%
  \BibitemOpen
  \bibfield  {author} {\bibinfo {author} {\bibfnamefont {M.}~\bibnamefont
  {Evans}}, \bibinfo {author} {\bibfnamefont {L.}~\bibnamefont {Barsotti}},\
  and\ \bibinfo {author} {\bibfnamefont {P.}~\bibnamefont {Fritschel}},\
  }\bibfield  {title} {\bibinfo {title} {A general approach to optomechanical
  parametric instabilities},\ }\href
  {https://doi.org/10.1016/j.physleta.2009.11.023} {\bibfield  {journal}
  {\bibinfo  {journal} {Phys. Lett. A}\ }\textbf {\bibinfo {volume} {374}},\
  \bibinfo {pages} {665} (\bibinfo {year} {2010})}\BibitemShut {NoStop}%
\bibitem [{\citenamefont {Chen}\ \emph {et~al.}(2015)\citenamefont {Chen},
  \citenamefont {Zhao}, \citenamefont {Danilishin}, \citenamefont {Ju},
  \citenamefont {Blair}, \citenamefont {Wang}, \citenamefont {Vyatchanin},
  \citenamefont {Molinelli}, \citenamefont {Kuhn}, \citenamefont {Gras},
  \citenamefont {Briant}, \citenamefont {Cohadon}, \citenamefont {Heidmann},
  \citenamefont {{Roch-Jeune}}, \citenamefont {Flaminio}, \citenamefont
  {Michel},\ and\ \citenamefont {Pinard}}]{Chen2015}%
  \BibitemOpen
  \bibfield  {author} {\bibinfo {author} {\bibfnamefont {X.}~\bibnamefont
  {Chen}}, \bibinfo {author} {\bibfnamefont {C.}~\bibnamefont {Zhao}}, \bibinfo
  {author} {\bibfnamefont {S.}~\bibnamefont {Danilishin}}, \bibinfo {author}
  {\bibfnamefont {L.}~\bibnamefont {Ju}}, \bibinfo {author} {\bibfnamefont
  {D.}~\bibnamefont {Blair}}, \bibinfo {author} {\bibfnamefont
  {H.}~\bibnamefont {Wang}}, \bibinfo {author} {\bibfnamefont {S.~P.}\
  \bibnamefont {Vyatchanin}}, \bibinfo {author} {\bibfnamefont
  {C.}~\bibnamefont {Molinelli}}, \bibinfo {author} {\bibfnamefont
  {A.}~\bibnamefont {Kuhn}}, \bibinfo {author} {\bibfnamefont {S.}~\bibnamefont
  {Gras}}, \bibinfo {author} {\bibfnamefont {T.}~\bibnamefont {Briant}},
  \bibinfo {author} {\bibfnamefont {P.-F.}\ \bibnamefont {Cohadon}}, \bibinfo
  {author} {\bibfnamefont {A.}~\bibnamefont {Heidmann}}, \bibinfo {author}
  {\bibfnamefont {I.}~\bibnamefont {{Roch-Jeune}}}, \bibinfo {author}
  {\bibfnamefont {R.}~\bibnamefont {Flaminio}}, \bibinfo {author}
  {\bibfnamefont {C.}~\bibnamefont {Michel}},\ and\ \bibinfo {author}
  {\bibfnamefont {L.}~\bibnamefont {Pinard}},\ }\bibfield  {title} {\bibinfo
  {title} {Observation of three-mode parametric instability},\ }\href
  {https://doi.org/10.1103/PhysRevA.91.033832} {\bibfield  {journal} {\bibinfo
  {journal} {Phys. Rev. A}\ }\textbf {\bibinfo {volume} {91}},\ \bibinfo
  {pages} {033832} (\bibinfo {year} {2015})}\BibitemShut {NoStop}%
\bibitem [{\citenamefont {Kharel}\ \emph {et~al.}(2019)\citenamefont {Kharel},
  \citenamefont {Harris}, \citenamefont {Kittlaus}, \citenamefont {Renninger},
  \citenamefont {Otterstrom}, \citenamefont {Harris},\ and\ \citenamefont
  {Rakich}}]{Kharel2019}%
  \BibitemOpen
  \bibfield  {author} {\bibinfo {author} {\bibfnamefont {P.}~\bibnamefont
  {Kharel}}, \bibinfo {author} {\bibfnamefont {G.~I.}\ \bibnamefont {Harris}},
  \bibinfo {author} {\bibfnamefont {E.~A.}\ \bibnamefont {Kittlaus}}, \bibinfo
  {author} {\bibfnamefont {W.~H.}\ \bibnamefont {Renninger}}, \bibinfo {author}
  {\bibfnamefont {N.~T.}\ \bibnamefont {Otterstrom}}, \bibinfo {author}
  {\bibfnamefont {J.~G.~E.}\ \bibnamefont {Harris}},\ and\ \bibinfo {author}
  {\bibfnamefont {P.~T.}\ \bibnamefont {Rakich}},\ }\bibfield  {title}
  {\bibinfo {title} {High-frequency cavity optomechanics using bulk acoustic
  phonons},\ }\href {https://doi.org/10.1126/sciadv.aav0582} {\bibfield
  {journal} {\bibinfo  {journal} {Sci. Adv.}\ }\textbf {\bibinfo {volume}
  {5}},\ \bibinfo {pages} {eaav0582} (\bibinfo {year} {2019})}\BibitemShut
  {NoStop}%
\bibitem [{\citenamefont {Evans}\ \emph {et~al.}(2015)\citenamefont {Evans}
  \emph {et~al.}}]{Evans2015}%
  \BibitemOpen
  \bibfield  {author} {\bibinfo {author} {\bibfnamefont {M.}~\bibnamefont
  {Evans}} \emph {et~al.},\ }\bibfield  {title} {\bibinfo {title} {Observation
  of parametric instability in {{Advanced LIGO}}},\ }\href
  {https://doi.org/10.1103/PhysRevLett.114.161102} {\bibfield  {journal}
  {\bibinfo  {journal} {Phys. Rev. Lett.}\ }\textbf {\bibinfo {volume} {114}},\
  \bibinfo {pages} {161102} (\bibinfo {year} {2015})}\BibitemShut {NoStop}%
\bibitem [{\citenamefont {Blair}\ \emph {et~al.}(2017)\citenamefont {Blair}
  \emph {et~al.}}]{Blair2017}%
  \BibitemOpen
  \bibfield  {author} {\bibinfo {author} {\bibfnamefont {C.}~\bibnamefont
  {Blair}} \emph {et~al.},\ }\bibfield  {title} {\bibinfo {title} {First
  demonstration of electrostatic damping of parametric instability at
  {{Advanced LIGO}}},\ }\href {https://doi.org/10.1103/PhysRevLett.118.151102}
  {\bibfield  {journal} {\bibinfo  {journal} {Phys. Rev. Lett.}\ }\textbf
  {\bibinfo {volume} {118}},\ \bibinfo {pages} {151102} (\bibinfo {year}
  {2017})}\BibitemShut {NoStop}%
\bibitem [{\citenamefont {Biscans}\ \emph {et~al.}(2019)\citenamefont
  {Biscans}, \citenamefont {Gras}, \citenamefont {Blair}, \citenamefont
  {Driggers}, \citenamefont {Evans}, \citenamefont {Fritschel}, \citenamefont
  {Hardwick},\ and\ \citenamefont {Mansell}}]{Biscans2019}%
  \BibitemOpen
  \bibfield  {author} {\bibinfo {author} {\bibfnamefont {S.}~\bibnamefont
  {Biscans}}, \bibinfo {author} {\bibfnamefont {S.}~\bibnamefont {Gras}},
  \bibinfo {author} {\bibfnamefont {C.~D.}\ \bibnamefont {Blair}}, \bibinfo
  {author} {\bibfnamefont {J.}~\bibnamefont {Driggers}}, \bibinfo {author}
  {\bibfnamefont {M.}~\bibnamefont {Evans}}, \bibinfo {author} {\bibfnamefont
  {P.}~\bibnamefont {Fritschel}}, \bibinfo {author} {\bibfnamefont
  {T.}~\bibnamefont {Hardwick}},\ and\ \bibinfo {author} {\bibfnamefont
  {G.}~\bibnamefont {Mansell}},\ }\bibfield  {title} {\bibinfo {title}
  {Suppressing parametric instabilities in {{LIGO}} using low-noise acoustic
  mode dampers},\ }\href {https://doi.org/10.1103/PhysRevD.100.122003}
  {\bibfield  {journal} {\bibinfo  {journal} {Phys. Rev. D}\ }\textbf {\bibinfo
  {volume} {100}},\ \bibinfo {pages} {122003} (\bibinfo {year}
  {2019})}\BibitemShut {NoStop}%
\bibitem [{\citenamefont {Hardwick}\ \emph {et~al.}(2020)\citenamefont
  {Hardwick}, \citenamefont {Hamedan}, \citenamefont {Blair}, \citenamefont
  {Green},\ and\ \citenamefont {{Vander-Hyde}}}]{Hardwick2020}%
  \BibitemOpen
  \bibfield  {author} {\bibinfo {author} {\bibfnamefont {T.}~\bibnamefont
  {Hardwick}}, \bibinfo {author} {\bibfnamefont {V.~J.}\ \bibnamefont
  {Hamedan}}, \bibinfo {author} {\bibfnamefont {C.}~\bibnamefont {Blair}},
  \bibinfo {author} {\bibfnamefont {A.~C.}\ \bibnamefont {Green}},\ and\
  \bibinfo {author} {\bibfnamefont {D.}~\bibnamefont {{Vander-Hyde}}},\
  }\bibfield  {title} {\bibinfo {title} {Demonstration of dynamic thermal
  compensation for parametric instability suppression in {{Advanced LIGO}}},\
  }\href {https://doi.org/10.1088/1361-6382/ab8be9} {\bibfield  {journal}
  {\bibinfo  {journal} {Classical Quantum Gravity}\ }\textbf {\bibinfo {volume}
  {37}},\ \bibinfo {pages} {205021} (\bibinfo {year} {2020})}\BibitemShut
  {NoStop}%
\bibitem [{\citenamefont {Aspelmeyer}\ \emph {et~al.}(2014)\citenamefont
  {Aspelmeyer}, \citenamefont {Kippenberg},\ and\ \citenamefont
  {Marquardt}}]{Aspelmeyer2014}%
  \BibitemOpen
  \bibfield  {author} {\bibinfo {author} {\bibfnamefont {M.}~\bibnamefont
  {Aspelmeyer}}, \bibinfo {author} {\bibfnamefont {T.~J.}\ \bibnamefont
  {Kippenberg}},\ and\ \bibinfo {author} {\bibfnamefont {F.}~\bibnamefont
  {Marquardt}},\ }\bibfield  {title} {\bibinfo {title} {Cavity optomechanics},\
  }\href {https://doi.org/10.1103/RevModPhys.86.1391} {\bibfield  {journal}
  {\bibinfo  {journal} {Rev. Mod. Phys.}\ }\textbf {\bibinfo {volume} {86}},\
  \bibinfo {pages} {1391} (\bibinfo {year} {2014})}\BibitemShut {NoStop}%
\bibitem [{\citenamefont {Corbitt}\ \emph {et~al.}(2006)\citenamefont
  {Corbitt}, \citenamefont {Ottaway}, \citenamefont {Innerhofer}, \citenamefont
  {Pelc},\ and\ \citenamefont {Mavalvala}}]{Corbitt2006}%
  \BibitemOpen
  \bibfield  {author} {\bibinfo {author} {\bibfnamefont {T.}~\bibnamefont
  {Corbitt}}, \bibinfo {author} {\bibfnamefont {D.}~\bibnamefont {Ottaway}},
  \bibinfo {author} {\bibfnamefont {E.}~\bibnamefont {Innerhofer}}, \bibinfo
  {author} {\bibfnamefont {J.}~\bibnamefont {Pelc}},\ and\ \bibinfo {author}
  {\bibfnamefont {N.}~\bibnamefont {Mavalvala}},\ }\bibfield  {title} {\bibinfo
  {title} {Measurement of radiation-pressure-induced optomechanical dynamics in
  a suspended {{Fabry-Perot}} cavity},\ }\href
  {https://doi.org/10.1103/PhysRevA.74.021802} {\bibfield  {journal} {\bibinfo
  {journal} {Phys. Rev. A}\ }\textbf {\bibinfo {volume} {74}},\ \bibinfo
  {pages} {021802(R)} (\bibinfo {year} {2006})}\BibitemShut {NoStop}%
\bibitem [{Note1()}]{Note1}%
  \BibitemOpen
  \bibinfo {note} {Corning ULE 7972, \SI {0.7}{\percent \per \cm } absorption
  at \SI {1064}{nm}.}\BibitemShut {Stop}%
\bibitem [{SM()}]{SM}%
  \BibitemOpen
  \href@noop {} {}\bibinfo {note} {See Supplemental Material, appended to this
  article, for an analysis of which spherical mirror geometries form stable
  acoustic cavities, the derivation of the parametric-instability power
  threshold and time dynamics, and closed-form expressions for the mode
  functions and overlap integrals, which includes
  Refs.~\cite{Yoshida1982,Sudhir2016PhDThesis,sdeint,Feldheim1940}}\BibitemShut
  {NoStop}%
\bibitem [{\citenamefont {Yoshida}(1982)}]{Yoshida1982}%
  \BibitemOpen
  \bibfield  {author} {\bibinfo {author} {\bibfnamefont {A.}~\bibnamefont
  {Yoshida}},\ }\bibfield  {title} {\bibinfo {title} {Spherical aberration in
  beam optical systems},\ }\href {https://doi.org/10.1364/AO.21.001812}
  {\bibfield  {journal} {\bibinfo  {journal} {Appl. Opt.}\ }\textbf {\bibinfo
  {volume} {21}},\ \bibinfo {pages} {1812} (\bibinfo {year}
  {1982})}\BibitemShut {NoStop}%
\bibitem [{\citenamefont {Sudhir}(2016)}]{Sudhir2016PhDThesis}%
  \BibitemOpen
  \bibfield  {author} {\bibinfo {author} {\bibfnamefont {V.}~\bibnamefont
  {Sudhir}},\ }\emph {\bibinfo {title} {{{Quantum limits on measurement and
  control of a mechanical oscillator}}}},\ \href
  {https://doi.org/10.5075/epfl-thesis-7202} {Ph.D. thesis},\ \bibinfo
  {school} {EPFL}, \bibinfo {address} {Lausanne} (\bibinfo {year}
  {2016})\BibitemShut {NoStop}%
\bibitem [{\citenamefont {Aburn}(2022)}]{sdeint}%
  \BibitemOpen
  \bibfield  {author} {\bibinfo {author} {\bibfnamefont {M.~J.}\ \bibnamefont
  {Aburn}},\ }\href {https://github.com/mattja/sdeint} {\bibinfo {title}
  {{{sdeint}} 0.3.0}} (\bibinfo {year} {2022}),\ \bibinfo {note}
  {\url{https://github.com/mattja/sdeint}}\BibitemShut {NoStop}%
\bibitem [{\citenamefont {Feldheim}(1940)}]{Feldheim1940}%
  \BibitemOpen
  \bibfield  {author} {\bibinfo {author} {\bibfnamefont {E.}~\bibnamefont
  {Feldheim}},\ }\bibfield  {title} {\bibinfo {title} {D\'eveloppements en
  s\'erie de polyn\^omes d'{{Hermite}} et de {{Laguerre}} \`a l'aide des
  transformations de {{Gauss}} et de {{Hankel I}}--{{III}}},\ }\href@noop {}
  {\bibfield  {journal} {\bibinfo  {journal} {Kon. Nederlandsche Akad. van Wet.
  Proc.}\ }\textbf {\bibinfo {volume} {43}},\ \bibinfo {pages} {224} (\bibinfo
  {year} {1940})}\BibitemShut {NoStop}%
\bibitem [{Not()}]{NoteBraginsky2001}%
  \BibitemOpen
  \href@noop {} {}\bibinfo {note} {In \cite{Braginsky2001}, the denominator in
  the parentheses in \cref{eq: def P threshold} is
  $(\gamma_1-\gamma_\mIndex)^2$, which we have corrected to
  $(\gamma_1+\gamma_\mIndex)^2$ (Supplemental Material~\cite{SM}). In the usual
  limit $\gamma_\mIndex\ll\gamma_1$, the results are equivalent}\BibitemShut
  {NoStop}%
\bibitem [{\citenamefont {Numata}\ \emph {et~al.}(2004)\citenamefont {Numata},
  \citenamefont {Kemery},\ and\ \citenamefont {Camp}}]{Numata2004}%
  \BibitemOpen
  \bibfield  {author} {\bibinfo {author} {\bibfnamefont {K.}~\bibnamefont
  {Numata}}, \bibinfo {author} {\bibfnamefont {A.}~\bibnamefont {Kemery}},\
  and\ \bibinfo {author} {\bibfnamefont {J.}~\bibnamefont {Camp}},\ }\bibfield
  {title} {\bibinfo {title} {Thermal-noise limit in the frequency stabilization
  of lasers with rigid cavities},\ }\href
  {https://doi.org/10.1103/PhysRevLett.93.250602} {\bibfield  {journal}
  {\bibinfo  {journal} {Phys. Rev. Lett.}\ }\textbf {\bibinfo {volume} {93}},\
  \bibinfo {pages} {250602} (\bibinfo {year} {2004})}\BibitemShut {NoStop}%
\bibitem [{\citenamefont {Saulson}(1990)}]{Saulson1990}%
  \BibitemOpen
  \bibfield  {author} {\bibinfo {author} {\bibfnamefont {P.~R.}\ \bibnamefont
  {Saulson}},\ }\bibfield  {title} {\bibinfo {title} {Thermal noise in
  mechanical experiments},\ }\href {https://doi.org/10.1103/PhysRevD.42.2437}
  {\bibfield  {journal} {\bibinfo  {journal} {Phys. Rev. D}\ }\textbf {\bibinfo
  {volume} {42}},\ \bibinfo {pages} {2437} (\bibinfo {year}
  {1990})}\BibitemShut {NoStop}%
\bibitem [{\citenamefont {Mason}\ and\ \citenamefont
  {Thurston}(1971)}]{MasonThurston1971}%
  \BibitemOpen
  \bibinfo {editor} {\bibfnamefont {W.~P.}\ \bibnamefont {Mason}}\ and\
  \bibinfo {editor} {\bibfnamefont {R.~N.}\ \bibnamefont {Thurston}},\ eds.,\
  \href@noop {} {\emph {\bibinfo {title} {Physical Acoustics: {{Principles}}
  and Methods}}},\ Vol.~\bibinfo {volume} {8}\ (\bibinfo  {publisher} {Academic
  Press},\ \bibinfo {address} {New York},\ \bibinfo {year} {1971})\BibitemShut
  {NoStop}%
\bibitem [{\citenamefont {Startin}\ \emph {et~al.}(1998)\citenamefont
  {Startin}, \citenamefont {Beilby},\ and\ \citenamefont
  {Saulson}}]{Startin1998}%
  \BibitemOpen
  \bibfield  {author} {\bibinfo {author} {\bibfnamefont {W.~J.}\ \bibnamefont
  {Startin}}, \bibinfo {author} {\bibfnamefont {M.~A.}\ \bibnamefont
  {Beilby}},\ and\ \bibinfo {author} {\bibfnamefont {P.~R.}\ \bibnamefont
  {Saulson}},\ }\bibfield  {title} {\bibinfo {title} {Mechanical quality
  factors of fused silica resonators},\ }\href
  {https://doi.org/10.1063/1.1149159} {\bibfield  {journal} {\bibinfo
  {journal} {Rev. Sci. Instrum.}\ }\textbf {\bibinfo {volume} {69}},\ \bibinfo
  {pages} {3681} (\bibinfo {year} {1998})}\BibitemShut {NoStop}%
\bibitem [{\citenamefont {Axelrod}(2024)}]{Axelrod2024Thesis}%
  \BibitemOpen
  \bibfield  {author} {\bibinfo {author} {\bibfnamefont {J.~J.}\ \bibnamefont
  {Axelrod}},\ }\emph {\bibinfo {title} {{{A laser phase plate for transmission
  electron microscopy}}}},\ \href {https://doi.org/10.48550/arXiv.2403.10670}
  {Ph.D. thesis},\ \bibinfo  {school} {University of California, Berkeley}
  (\bibinfo {year} {2024}),\ \Eprint {https://arxiv.org/abs/2403.10670}
  {arXiv:2403.10670} \BibitemShut {NoStop}%
\bibitem [{Note2()}]{Note2}%
  \BibitemOpen
  \bibinfo {note} {Schott Zerodur Expansion Class 0 Special.}\BibitemShut
  {Stop}%
\bibitem [{\citenamefont {Maisenbacher}\ \emph {et~al.}(2026)\citenamefont
  {Maisenbacher}, \citenamefont {Singh}, \citenamefont {Pope},\ and\
  \citenamefont {M{\"u}ller}}]{DataPI2026}%
  \BibitemOpen
  \bibfield  {author} {\bibinfo {author} {\bibfnamefont {L.}~\bibnamefont
  {Maisenbacher}}, \bibinfo {author} {\bibfnamefont {A.}~\bibnamefont {Singh}},
  \bibinfo {author} {\bibfnamefont {I.~M.}\ \bibnamefont {Pope}},\ and\
  \bibinfo {author} {\bibfnamefont {H.}~\bibnamefont {M{\"u}ller}},\ }\href
  {https://doi.org/10.5281/zenodo.23692011} {\bibinfo {title} {Data for
  ``{{Optomechanical}} limit on intracavity intensities from bulk mirror
  vibrations''}},\ \bibinfo {howpublished} {Version 1, [Dataset], Zenodo}
  (\bibinfo {year} {2026}),\ \bibinfo {note}
  {\url{https://doi.org/10.5281/zenodo.23692011}}\BibitemShut {NoStop}%
\bibitem [{\citenamefont {Kharel}\ \emph {et~al.}(2018)\citenamefont {Kharel},
  \citenamefont {Chu}, \citenamefont {Power}, \citenamefont {Renninger},
  \citenamefont {Schoelkopf},\ and\ \citenamefont {Rakich}}]{Kharel2018}%
  \BibitemOpen
  \bibfield  {author} {\bibinfo {author} {\bibfnamefont {P.}~\bibnamefont
  {Kharel}}, \bibinfo {author} {\bibfnamefont {Y.}~\bibnamefont {Chu}},
  \bibinfo {author} {\bibfnamefont {M.}~\bibnamefont {Power}}, \bibinfo
  {author} {\bibfnamefont {W.~H.}\ \bibnamefont {Renninger}}, \bibinfo {author}
  {\bibfnamefont {R.~J.}\ \bibnamefont {Schoelkopf}},\ and\ \bibinfo {author}
  {\bibfnamefont {P.~T.}\ \bibnamefont {Rakich}},\ }\bibfield  {title}
  {\bibinfo {title} {Ultra-high-{{Q}} phononic resonators on-chip at cryogenic
  temperatures},\ }\href {https://doi.org/10.1063/1.5026798} {\bibfield
  {journal} {\bibinfo  {journal} {APL Photonics}\ }\textbf {\bibinfo {volume}
  {3}},\ \bibinfo {pages} {066101} (\bibinfo {year} {2018})}\BibitemShut
  {NoStop}%
\bibitem [{\citenamefont {Renninger}\ \emph {et~al.}(2018)\citenamefont
  {Renninger}, \citenamefont {Kharel}, \citenamefont {Behunin},\ and\
  \citenamefont {Rakich}}]{Renninger2018}%
  \BibitemOpen
  \bibfield  {author} {\bibinfo {author} {\bibfnamefont {W.~H.}\ \bibnamefont
  {Renninger}}, \bibinfo {author} {\bibfnamefont {P.}~\bibnamefont {Kharel}},
  \bibinfo {author} {\bibfnamefont {R.~O.}\ \bibnamefont {Behunin}},\ and\
  \bibinfo {author} {\bibfnamefont {P.~T.}\ \bibnamefont {Rakich}},\ }\bibfield
   {title} {\bibinfo {title} {Bulk crystalline optomechanics},\ }\href
  {https://doi.org/10.1038/s41567-018-0090-3} {\bibfield  {journal} {\bibinfo
  {journal} {Nat. Phys.}\ }\textbf {\bibinfo {volume} {14}},\ \bibinfo {pages}
  {601} (\bibinfo {year} {2018})}\BibitemShut {NoStop}%
\bibitem [{\citenamefont {Lay}\ and\ \citenamefont
  {Wallace}(1995)}]{LayWallace1995}%
  \BibitemOpen
  \bibfield  {author} {\bibinfo {author} {\bibfnamefont {T.}~\bibnamefont
  {Lay}}\ and\ \bibinfo {author} {\bibfnamefont {T.~C.}\ \bibnamefont
  {Wallace}},\ }\href@noop {} {\emph {\bibinfo {title} {Modern Global
  Seismology}}},\ \bibinfo {series} {International {{Geophysics Series}}}\
  No.~\bibinfo {number} {58}\ (\bibinfo  {publisher} {Academic Press},\
  \bibinfo {address} {San Diego},\ \bibinfo {year} {1995})\BibitemShut
  {NoStop}%
\bibitem [{\citenamefont {Cook}\ and\ \citenamefont
  {Arnoult}(1976)}]{Cook1976}%
  \BibitemOpen
  \bibfield  {author} {\bibinfo {author} {\bibfnamefont {B.~D.}\ \bibnamefont
  {Cook}}\ and\ \bibinfo {author} {\bibfnamefont {W.~J.}\ \bibnamefont
  {Arnoult}, \bibfnamefont {III}},\ }\bibfield  {title} {\bibinfo {title}
  {Gaussian--{{Laguerre}}/{{Hermite}} formulation for the nearfield of an
  ultrasonic transducer},\ }\href {https://doi.org/10.1121/1.380832} {\bibfield
   {journal} {\bibinfo  {journal} {J. Acoust. Soc. Am.}\ }\textbf {\bibinfo
  {volume} {59}},\ \bibinfo {pages} {9} (\bibinfo {year} {1976})}\BibitemShut
  {NoStop}%
\bibitem [{\citenamefont {Newberry}\ and\ \citenamefont
  {Thompson}(1989)}]{Newberry1989}%
  \BibitemOpen
  \bibfield  {author} {\bibinfo {author} {\bibfnamefont {B.~P.}\ \bibnamefont
  {Newberry}}\ and\ \bibinfo {author} {\bibfnamefont {R.~B.}\ \bibnamefont
  {Thompson}},\ }\bibfield  {title} {\bibinfo {title} {A paraxial theory for
  the propagation of ultrasonic beams in anisotropic solids},\ }\href
  {https://doi.org/10.1121/1.397775} {\bibfield  {journal} {\bibinfo  {journal}
  {J. Acoust. Soc. Am.}\ }\textbf {\bibinfo {volume} {85}},\ \bibinfo {pages}
  {2290} (\bibinfo {year} {1989})}\BibitemShut {NoStop}%
\bibitem [{\citenamefont {Lax}\ \emph {et~al.}(1975)\citenamefont {Lax},
  \citenamefont {Louisell},\ and\ \citenamefont {McKnight}}]{Lax1975}%
  \BibitemOpen
  \bibfield  {author} {\bibinfo {author} {\bibfnamefont {M.}~\bibnamefont
  {Lax}}, \bibinfo {author} {\bibfnamefont {W.~H.}\ \bibnamefont {Louisell}},\
  and\ \bibinfo {author} {\bibfnamefont {W.~B.}\ \bibnamefont {McKnight}},\
  }\bibfield  {title} {\bibinfo {title} {From {{Maxwell}} to paraxial wave
  optics},\ }\href {https://doi.org/10.1103/PhysRevA.11.1365} {\bibfield
  {journal} {\bibinfo  {journal} {Phys. Rev. A}\ }\textbf {\bibinfo {volume}
  {11}},\ \bibinfo {pages} {1365} (\bibinfo {year} {1975})}\BibitemShut
  {NoStop}%
\bibitem [{\citenamefont {Rasband}(1975)}]{Rasband1975}%
  \BibitemOpen
  \bibfield  {author} {\bibinfo {author} {\bibfnamefont {S.~N.}\ \bibnamefont
  {Rasband}},\ }\bibfield  {title} {\bibinfo {title} {Resonant vibrations of
  free cylinders and disks},\ }\href {https://doi.org/10.1121/1.380531}
  {\bibfield  {journal} {\bibinfo  {journal} {J. Acoust. Soc. Am.}\ }\textbf
  {\bibinfo {volume} {57}},\ \bibinfo {pages} {899} (\bibinfo {year}
  {1975})}\BibitemShut {NoStop}%
\end{thebibliography}
\end{document}


\pagestyle{plain}

\phantomsection\label{supplemental}

\widetext
\begin{center}
\textbf{\large Supplemental Material for \textit{Optomechanical Limit on Intracavity Intensities from Bulk Mirror Vibrations}}
\end{center}

\phantomeq{1}{eq:DressedDecayTime}
\phantomeq{2}{eq: def P threshold}
\phantomeq{3}{eq: def B}
\phantomfig{1}{fig:MirrorModeFigure}
\phantomfig{2}{fig:Signals}

\section{\label{SM:mirror geometry}Choice of mirror geometry}

\begin{figure}[b]
    \includegraphics{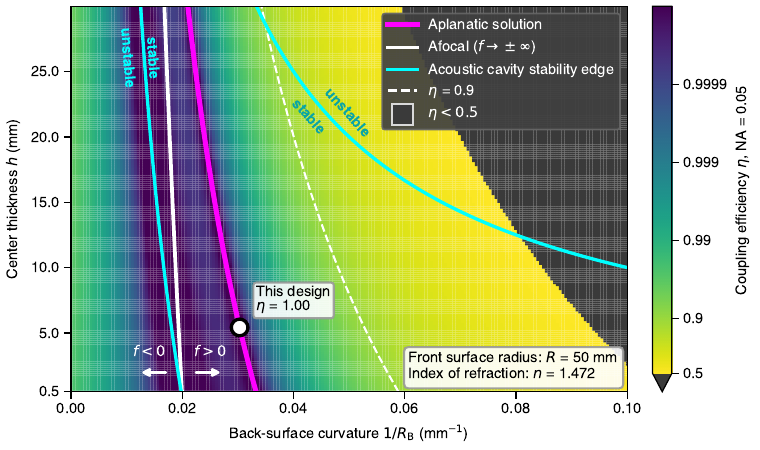}
    \caption{\label{sm:fig:SphericalAberrationsR50}
    Comparison of different spherical cavity mirror geometries for a concave front surface with $\RS = \SI{50}{mm}$ radius of curvature.
    The color indicates the coupling efficiency $\eta$ of the input beam (i.e., a beam entering from the back surface and transmitted through the mirror) into the cavity mode for a cavity $\mathrm{NA}$ of \num{0.05}.
    $\eta$ is limited by spherical aberrations.
    It is calculated by first ray tracing a point source from the cavity center through to the plane of the mirror's back surface vertex.
    The optical path length of the rays gives the geometric optical wavefront, which is compared to the unaberrated Gaussian beam's phase front, and the difference is interpreted as a wavefront error $\phi_\mathrm{B}$.
    By multiplying the unaberrated Gaussian beam with a phase mask $e^{ik\phi_\mathrm{B}}$ ($k$: optical wavenumber), we generate the aberrated Gaussian beam \cite{Yoshida1982}.
    $\eta$ is the squared modulus of the overlap integral of this aberrated Gaussian beam with an input Gaussian beam, whose waist size and position are optimized to maximize $\eta$.
    This method is accurate in the limit of small aberrations ($k\phi_\mathrm{B}<1$ across the Gaussian beam diameter).
    Cyan lines mark the region in which the mirror forms a stable acoustic cavity; the white line marks the boundary between the mirror forming a converging ($f > 0$) or diverging ($f < 0$) lens; the pink line marks the aplanatic solution, which is free of leading-order spherical aberrations (all three meet at $\mirrorthickness = n \RS$).
    All geometries with $f > 0$ and sufficiently high coupling efficiency $\eta > 0.9$ (and $\mirrorthickness \lesssim \SI{30}{mm}$) form a stable acoustic cavity.
    The design used here (white circle) is aplanatic with $\RBS = \SI{33}{mm}$ and $\mirrorthickness = \SI{5.4}{mm}$ (\cref{fig:MirrorModeFigure}).
    A suitable geometry with an unstable acoustic cavity would be, e.g., $\RBS \approx \SI{65}{mm}$ and $\mirrorthickness \lesssim \SI{10}{mm}$ combined with an additional low-aberration coupling lens whose geometry is not constrained by an acoustic cavity condition.
    The index of refraction $n = 1.472$ of ULE at \SI{1064}{nm} is used.
    }
\end{figure}

\begin{figure}[t]
    \includegraphics{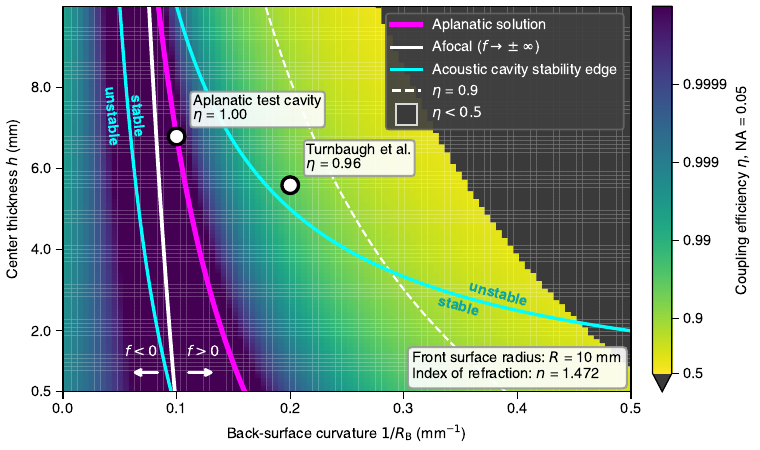}
    \caption{\label{sm:fig:SphericalAberrationsR10}
    Same as \cref{sm:fig:SphericalAberrationsR50}, but for $\RS = \SI{10}{mm}$.
    }
\end{figure}

A natural choice for the mirror geometry of a high-NA cavity is a geometry that focuses a collimated input beam such that it overlaps with the cavity mode.
To ensure high overlap, the geometry should minimize aberrations, in particular spherical aberrations, which is achieved by a convex-concave mirror design that minimizes the angles of incidence.
Such a design tends to form a stable acoustic cavity.
This situation is illustrated in \cref{sm:fig:SphericalAberrationsR50} for possible spherical mirror geometries with a concave (front) surface with $\RS = \SI{50}{mm}$ radius of curvature and for a cavity NA of \num{0.05}, as used in this work.
We find that all geometries that form a converging lens ($f > 0$) and possess sufficiently low aberrations (and thickness $\mirrorthickness \lesssim \SI{30}{mm}$) form a stable acoustic cavity.
This includes the mirror design used in this work (\cref{fig:MirrorModeFigure} of the main text), which is an aplanatic design that is free of leading-order spherical aberrations \cite{Axelrod2024Thesis}.
To avoid forming a stable acoustic cavity, a mirror geometry forming a diverging lens ($f < 0$) must be used, in combination with a separate low-aberration coupling lens that generates the high-NA input beam.
Importantly, the design of the coupling lens, unlike for the cavity mirrors, is not constrained by an acoustic cavity condition.
We here use the index of refraction $n = 1.472$ of ULE at \SI{1064}{nm}; using instead $n = 1.53$ for Zerodur at the same wavelength does not change the results substantially.

For comparison, \cref{sm:fig:SphericalAberrationsR10} shows the situation for $\RS = \SI{10}{mm}$, which matches the cavity mirrors used in \cite{Turnbaugh2021,Axelrod2024Thesis} and in the short test cavity mentioned in the main text.
Here, solutions with $f > 0$ and sufficiently low aberrations that do not form a stable acoustic cavity can be found, including the geometry of \cite{Turnbaugh2021} that was used to demonstrate intensities up to \SI{450}{GW/cm^2} at $\mathrm{NA} \approx \num{0.04}$.
However, aplanatic designs again form a stable acoustic cavity; such a design~\cite{Axelrod2024Thesis} was used in the short test cavity in which we believe we observed PI.

Significant aberrations can also occur when the axis of symmetry of a mirror's back surface is not aligned with the cavity axis, which can result from a tilt of the cavity mode with respect to the mirror or from manufacturing tolerances.
The resulting optimal coupling efficiencies can be much lower than suggested by \cref{sm:fig:SphericalAberrationsR50,sm:fig:SphericalAberrationsR10}.
However, designs that minimize spherical aberrations also tend to minimize the effect of such misalignments on the cavity coupling, which makes the spherical aberrations studied here a good figure of merit for the choice of mirror geometry.

\section{\label{SM:DetailsofPIThreshold}Details of the PI threshold and time dynamics}

Using the same methods as~\cite{Braginsky2001}, we can derive dynamical equations of motion for the slowly varying field amplitudes $D_0$, $D_1$, and $X$ of the $\TEMfundamental$, $\TEMHOM$ and mechanical modes (we here omit the mode index $j$), respectively, in the rotating-wave approximation:
\begin{subequations}
\begin{align}
    \dot D_0+\gamma_0(D_0-D_{0,\mathrm{sp}}) = -i\omega_1B^*XD_1e^{i\Delta\omega t}/L,\label{eq: D0 dot}\\
\dot D_1+\gamma_1D_1 =-i\omega_0BX^*D_0e^{-i\Delta\omega t}/L,\label{eq: D1 dot}\\
\dot X + \gamma_\m X = \frac{-i\epsilon_0\omega_0\omega_1}{2M\omega_\m}BD_0D_1^*e^{-i\Delta\omega t}+\hat f_n\label{eq: X dot},
\end{align}
\end{subequations}
where we have added a drive field (`setpoint') $D_{0,\mathrm{sp}}$ and a stochastic noise term $\hat f_n$ to the differential equations of~\cite{Braginsky2001} ($\gamma_0$: amplitude decay constant of $\TEMfundamental$ mode).
The exact form of $\hat f_n$ is unimportant for the overall dynamics, as its sole purpose is to provide an initial excitation of the mechanical mode.
We choose a Langevin noise term $\hat f_n = \sqrt{\gamma_\m \kB T/(2M\omega_\m^2)}R(t)$, where $R(t)$ is a delta-correlated Gaussian white noise process with a random complex phase ($\kB$: Boltzmann constant).
With this definition of slowly varying amplitudes, the circulating electric field is
\begin{equation}
    F_\mathrm{circ}(\bm r, t) = \sum_{a=0,1} \Re\left(i\omega_aD_a(t)U_{a0}(\bm r)e^{i(k_{a}z-\omega_{a} t)}\right),
\end{equation}
with the Hermite-Gauss envelopes $U_{a0}$ defined in the following section.
The circulating power in an optical cavity mode indexed by $a\in\{0,1\}$ is $P_{\mathrm{circ},a} = \epsilon_0c \omega_a^2|D_a(t)|^2/2,$
while the stored energy in the cavity mode is $E_a = 2LP_{\mathrm{circ},a}/c$, and the stored energy in the mechanical mode is $E_\m = 2M\omega_\m^2|X|^2$.

\Cref{eq: D0 dot,eq: D1 dot,eq: X dot} apply to the here-relevant case, where the optomechanical coupling mechanism is radiation pressure acting on the mirror surface.
Electrostrictive optomechanical coupling in the bulk of a mechanical resonator has also been observed to induce PI \cite{Kharel2019}.
However, in our optical cavity, the region of highest electric field is localized within a few layers of the mirror coating, where the acoustic mode has nearly zero strain due to its boundary condition.
The electrostrictive interaction is thus orders of magnitude weaker than radiation pressure, and can be ignored.
Thermoelastic optomechanical coupling, as seen in \cite{Sudhir2016PhDThesis}, is similarly orders of magnitude too small to consider in this work, as the PI~frequencies of several \si{MHz} are much faster than the thermal diffusion rates in the coating ($\sim$\si{kHz}), which itself has a small coefficient of thermal expansion (\SI{\sim d-6}{K^{-1}}).

\begin{figure}
    \centering
    \includegraphics{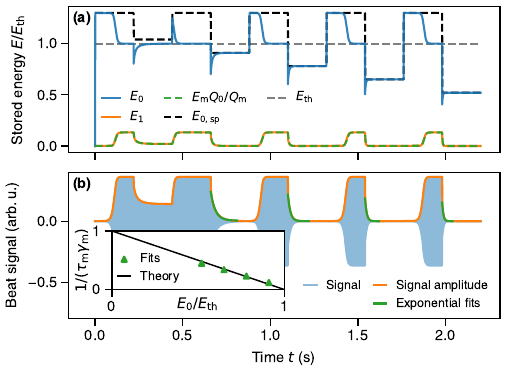}
    \caption{\label{fig: SDE int}
    Numerical integration of \cref{eq: D0 dot,eq: D1 dot,eq: X dot}.
    (a)~Stored energy in the $\TEMfundamental$, $\TEMHOM$ and mechanical modes ($E_0$, $E_1$ and $E_\m$, respectively), relative to the PI threshold energy $E_\mathrm{th}=2L\Pthresh/c$.
    Due to the drastically different scales of stored mechanical and optical energy, we show $E_\m$ multiplied by $Q_0/Q_\m$, which has the same (average) magnitude as $E_1$ ($Q_0$, $Q_\m$: quality factors of the $\TEMfundamental$ and mechanical modes).
    $E_0$ initially follows the drive value $E_{0,\mathrm{sp}}\propto|D_{0,\mathrm{sp}}|^2$ (black dashed line), before a spontaneous thermal fluctuation from $\hat f_n$ initiates oscillation through PI, and $E_0$ drops to the threshold value (gray dashed line).
    The setpoint energy is stepped down in the same manner as in the experiment (\cref{fig:Signals} of the main text).
    In the simulation, $Q_0=\num{7e5}\,Q_\m$, $\gamma_1 = \gamma_0$, $\gamma_\m = \gamma_0/400$, $\Delta\omega = -\gamma_0$, and $\gamma_0 = 2\pi\times\SI{20}{kHz}$, which are typical values.
    (b)~The beat signal $D_0D_1^*$ (in the rotating frame) oscillates at $\sigma \approx -\Delta\omega$ (unresolved on this time scale; blue band); its amplitude $|D_0D_1^*|$ is shown in orange.
    At each step-down of the drive field to a value below the threshold, an exponential decay is fit to the amplitude (green lines).
    The inset shows the time constants of the fits against the circulating power after the step-down, in good agreement with the theoretical prediction of \cref{eq:DressedDecayTime} of the main text.
    }
\end{figure}

The system of nonlinear, stochastic differential equations \cref{eq: D0 dot,eq: D1 dot,eq: X dot} is integrated numerically~\cite{sdeint}, and the results are shown in \cref{fig: SDE int}.
When the $\TEMfundamental$ power is driven above the threshold, after a short time, its power drops, while the power circulating in the $\TEMHOM$ and mechanical modes grows.
$\Pcirc$ settles to the threshold value given in \cref{eq: def P threshold} of the main text, where it remains constant.
The dynamics in the simulation qualitatively agree with the dynamics in \cref{fig:Signals}, though the specifics depend on the mode being measured.

To derive the power threshold, we assume the $D_0$ field attains a steady-state value, neglect the small noise term $\hat f_n$, and look at the reduced system of differential equations for $D_1$ and $X$, which is now linear. Introducing $\xi = Xe^{i\Delta\omega t}$, we find a time-independent linear system in $D_1$ and $\xi^*$, whose characteristic equation is
\begin{equation}\label{eq: characteristic equation}
    (-\gamma_1-\lambda)(-\gamma_\m-i\Delta\omega-\lambda)-A = 0,
\end{equation}
where $A = \epsilon_0\omega_0^2\omega_1|B|^2|D_0|^2/(2M\omega_\m L)$.
The power threshold is given by the value for $D_0$ above which the eigenvalue $\lambda_+$ has a positive real part, where
\begin{equation}\label{eq: lambda plus}
    \lambda_+ = \frac{-\gamma_1-\gamma_\m-i\Delta\omega+\sqrt{(\gamma_1-\gamma_\m-i\Delta\omega)^2+4A}}{2}.
\end{equation}

At this point, \cite{Braginsky2001} makes an approximation in smallness of $A$ to show the existence of the power threshold, and computes it within this approximation. However, knowing a solution exists, we can write it down exactly by asserting that at the power threshold, one eigenvalue becomes $\lambda = 0+i\sigma$, where $\sigma\in\mathbb{R}$.
By equating the imaginary parts of \cref{eq: characteristic equation}, we find
\begin{equation}
    \sigma = \frac{-\gamma_1}{\gamma_1+\gamma_\m}\Delta\omega\approx -\Delta\omega.
\end{equation}
Thus, during PI, $D_1\sim e^{i\sigma t}$, which tells us the oscillation frequency of the PI~signal from
\begin{equation}
    D_0D_1^*e^{-i(\omega_0-\omega_1)t}\sim e^{-i(\omega_0-\omega_1+\sigma)t}\approx e^{-i\omega_\m t},
\end{equation}
i.e., $\PIFreq\approx\omega_\m/(2\pi)$, regardless of $\Delta\omega$.

Meanwhile, by equating the real parts of \cref{eq: characteristic equation}, we can compute the threshold
\begin{equation}
    \Athresh = \gamma_1\gamma_\m\left(1+\frac{\Delta\omega^2}{(\gamma_1+\gamma_\m)^2}\right),
\end{equation}
which rearranges to \cref{eq: def P threshold} of the main text.
Note that since $\gamma_\m\ll\gamma_1$, we have $A\ll\gamma_1^2+\Delta\omega^2$ for powers around or below the power threshold, which justifies the approximation at the end of the Appendix of~\cite{Braginsky2001}.
Within this approximation, we can use \cref{eq: lambda plus} to write the mechanical mode exponential decay time $\tau_\m=-1/\Re(\lambda_+)$ in the form of \cref{eq:DressedDecayTime}.

This work assumes each PI only involves one mechanical mode at a time. While in general, many mechanical modes can be coherently driven by the beating optical fields, in practice the extremely narrow mechanical mode linewidths lead to only one mode being excited in any particular instance of PI. Multiple distinct PIs can be driven simultaneously, however, if $\Pcirc$ exceeds the threshold of PI for two different mechanical modes, which can happen in the transient period before onset of PI above its power threshold.

\section{\label{SM:explicit forms}Explicit forms for mode functions and overlap integrals}

In this section, we need to be careful to distinguish between properties of the optical modes and of the mechanical modes, which both have similar forms. We will use the convention that a tilde denotes a mechanical property where it may otherwise be confused with an optical property. This is in contrast to the notation used in the main text and the preceding section where the subscript `m' was used to label mechanical properties, as we here need subscripts to index several other properties.

The paraxial wave equation, which may describe either optical or mechanical modes in this work, is given by
\begin{equation}\label{eq: supplementary paraxial wave equation}
    \nabla_\perp^2U+2i\kanon\partial_zU=0.
\end{equation}
The Hermite-Gaussian modes, which are traveling-wave solutions to \cref{eq: supplementary paraxial wave equation}, are parametrized by the beam waist $\wanon$ and the wavenumber $\kanon$ and indexed by transverse indices $m$ and $n$. They are
\begin{equation}\label{eq: HGM mode profile}
    U_{mn}(\bm r;\kanon,\wanon,\zanon) = \frac{1}{\sqrt{2^{m+n-1}m!n!\pi \wanonz^2}}\Hermite_m\left(\frac{\sqrt 2 x}{\wanonz}\right)\Hermite_n\left(\frac{\sqrt 2 y}{\wanonz}\right)e^{-\frac{x^2+y^2}{\wanonz^2}}e^{\frac{ik(x^2+y^2)}{2R(z)}}e^{-i\left(m+n+1\right)\psi(z)},
\end{equation}
where $\wanonz = \wanon\sqrt{1+(z-\zanon)^2/z_R^2}$, $z_R = k(\wanon)^2/2$, $R(z) = (z-\zanon)+z_R^2/(z-\zanon)$, $\psi(z) = \tan^{-1}((z-\zanon)/z_R)$, $\Hermite_m(x)$ is the Hermite polynomial of degree $m$, and $\zanon$ is the position of the beam's waist along the propagation ($z$) axis.
These functions are normalized so that $\int U_{mn}(\bm r_\perp,z)U_{m'n'}^*(\bm r_\perp,z)d\bm r_\perp = \delta_{mm'}\delta_{nn'}$, where $\delta$ is the Kronecker delta.

The mechanical (scalar potential) mode functions $\Phi_j$ are indexed by $j$, which refers to the tuple of indices $j = (m_j,n_j,p_j)$ ($m_j$: $x$-transverse mode index; $n_j$: $y$-transverse mode index; $p_j$: longitudinal mode index). They are given by
\begin{equation}
    \Phi_{j}(\bm r, t) =A_{j}\Re\left[\Im\left[e^{i\varphi_{j}}U_{m_jn_j}(\bm r;\tilde k_{j}, \wmechAppendix)e^{i\tilde k_{j}z}\right]e^{-i\tilde \omega_{j} t}\right],
\end{equation}
where $\wmechAppendix$ is the mechanical mode waist and $\varphi_j = (m_j+n_j+1)\tilde \psi(\zs)-\tilde k_{j}\,\zs$, with $\zs$ the position of the mirror's front (reflective) surface.
$\tilde \omega_{j} = c_L\tilde k_{j}$ is the mechanical mode angular frequency,
\begin{equation}\label{eq: mode frequencies}
    \frac{\tilde \omega_{j}}{2\pi} = \frac{c_L}{2\mirrorthickness}\left(p_j+\frac{(m_j+n_j+1)}{\pi}\cos^{-1}\left(\sqrt{\tilde g_1\tilde g_2}\right)\right),
\end{equation}
where $\tilde g_1 = 1-\mirrorthickness/\RBS=0.84$ and $\tilde g_2 = 1-\mirrorthickness/(-\RS)=1.11$ are the mechanical cavity g-parameters, and $\mirrorthickness$ is the mirror thickness. The amplitude $A_{j}$ is an arbitrary choice of normalization that is canceled out in $|B|^2/M$: we choose the convention $A_{j} = \sqrt{2\pi}\wmechAppendix/\tilde k_{j}$, which makes $B$ dimensionless and $M$ have dimensions of mass. The displacement amplitude $u_z$ that appears in \cref*{eq: def B} of the main text is, at leading order in the paraxial parameter and up to appropriate factors of $i$,
\begin{equation}
    u_{z, j}(\bm r_\perp) = \sqrt{2\pi}\wmechAppendix U_{m_jn_j}(x, y, \zs; \tilde k_{j}, \wmechAppendix, \zzeromech).
\end{equation}
The mirror front surface position $\zs$ relative to the mechanical mode waist at $\zzeromech$ and the mechanical Rayleigh range are given by
\begin{align}
    \zs - \zzeromech &= \mirrorthickness-\frac{\mirrorthickness\tilde g_2(1-\tilde g_1)}{\tilde g_1+\tilde g_2-2\tilde g_1\tilde g_2} = \SI{-5.37}{mm},\\
    \tilde z_R &= \sqrt{\frac{\tilde g_1\tilde g_2(1-\tilde g_1\tilde g_2)\mirrorthickness^2}{(2\tilde g_1\tilde g_2-\tilde g_1-\tilde g_2)^2}} = \SI{15.48}{mm},
\end{align}
from which $\wmechAppendix = \sqrt{2\tilde z_R/\tilde k_{j}}$ follows. The effective masses of the modes are
\begin{equation}
    M_{j} = \pi \wmechAppendix^2\mirrorthickness\rho,
\end{equation}
where $\rho$ is the mirror density.

The optical mode functions in \cref*{eq: def B} of the $\TEMfundamental$ and $\TEMHOM$ modes are, up to appropriate factors of $i$,
\begin{align}
   & F_0(\bm r_\perp) = U_{00}(\bm r_\perp, \zs; \omega_0/c, \wzeroopt,  \zzeroopt),
   & F_1(\bm r_\perp) = U_{10}(\bm r_\perp, \zs; \omega_1/c, \wzeroopt, \zzeroopt),
\end{align}
where $\wzeroopt$ is the optical mode waist and $\zs - \zzeroopt \approx L/2$ is the position of the mirror front surface relative to the optical waist.

Computing the overlap integral $B$ in \cref*{eq: def B} requires an integral over the curved spherical surface.
However, within the paraxial approximation, we can compute the overlap integral over a constant $z$ plane defined by the center of the mirror's curved surface (the surface $S$), provided we ignore the wavefront curvature factor in each $U_{mn}$ function (which is matched to the spherical surface curvature).
In the general case, the principal axes of the mechanical modes may be rotated by an angle $\theta$ relative to those of the optical modes, which can in turn be offset from the mechanical mode axes by $(x_c, y_c)$.
In this context, the overlap integrals are (up to a global phase)

\begin{equation}
B_{j}(x_c, y_c, \theta) = C_j\int_{\mathbb{R}^2} (x-x_c)\Hermite_{m_j}\left(\frac{\sqrt 2 x'}{\wzsmech}\right)\Hermite_{n_j}\left(\frac{\sqrt 2 y'}{\wzsmech}\right)\exp\left(-\frac{x^2+y^2}{\wzsmech^2}-2\frac{(x-x_c)^2+(y-y_c)^2}{\wzsopt^2}\right) dx\,dy,
\end{equation}
where
\begin{equation}
    C_j = \frac{8\wmechAppendix}{\pi\wzsopt^3\wzsmech\sqrt{2^{(m_j+n_j)}m_j!n_j!}}.
\end{equation}
$\wzsopt$ and $\wzsmech$ are the optical and mechanical beam radii at the mirror front surface $S$, respectively, and $(x', y') = (x\cos\theta+y\sin\theta, -x\sin\theta+y\cos\theta)$. This formula has a closed-form result, which can be computed using Feldheim's formulae (46) and (77) \cite{Feldheim1940} (and correcting an erroneous $(-1)^k$ which should read $(-1)^r$ in Feldheim's (77)).

\begin{equation}
    B_{j}(x_c,y_c,\theta) = \sum_{l=0}^{m_j+n_j}\sqrt{\frac{l!(m_j+n_j-l)!}{m_j!n_j!}}a_{m_jn_j;l}(\theta)b_{l,m_j+n_j-l},
\end{equation}
where
\begin{align}
&a_{\mu\nu;l}(\theta) = \sum_{r=0}^l(-1)^{r}\binom{\nu}{r}\binom{\mu}{l-r}\sin^{\mu+2r-l}(\theta)\cos^{\nu+l-2r}(\theta),\\
&b_{\mu\nu} = \frac{8\wmechAppendix W_e^4\left(1-2W_{e}^2/\wzsmech^2\right)^{\frac{\mu+\nu}{2}}}{\wzsopt^3\wzsmech\sqrt{2^{\mu+\nu}\mu!\nu!}}\Hermite_\nu\left(A_y\right)\left[\frac{\mu\sqrt 2 \Hermite_{\mu-1}\left(A_x\right)}{\sqrt{\wzsmech^2-2W_{e}^2}}-\frac{x_c\Hermite_\mu\left(A_x\right)}{\wzsmech^2} \right]\exp\left({\frac{-2(x_c^2+y_c^2)}{\wzsopt^2+2\wzsmech^2}}\right),
\end{align}
and
\begin{align}
&W_e=\sqrt{\frac{\wzsmech^2\wzsopt^2}{2\wzsmech^2+\wzsopt^2}},&
&A_x = \frac{2\sqrt 2 x_c W_{e}^2}{\wzsopt^2\sqrt{\wzsmech^2-2W_{e}^2}},&
&A_y = \frac{2\sqrt 2 y_c W_{e}^2}{\wzsopt^2\sqrt{\wzsmech^2-2W_{e}^2}}.&
\end{align}
In particular, in the case that the optical and mechanical modes are perfectly aligned ($x_c=y_c=0$ and $\theta=0$), we find
\begin{align}
B_{j}(0,0,0) = \begin{cases}\frac{8\wmechAppendix}{\sqrt{m_j!n_j!}}\frac{\wzsopt\wzsmech^2(\wzsopt^2-2\wzsmech^2)^{(m_j+n_j-1)/2}}{(\wzsopt^2+2\wzsmech^2)^{(m_j+n_j+3)/2}}m_j!!(n_j-1)!!  &(\mathrm{if}\,m_j\, \mathrm{odd}\,\mathrm{and}\,n_j\,\mathrm{even}),\\
0&(\mathrm{otherwise}),
\end{cases}
\end{align}
where $!!$ denotes the double factorial, with the special case that $(-1)!!$ is interpreted to mean 1.